\documentclass[reprint, amsmath,amssymb, aps]{revtex4-1}
\usepackage{graphicx}
\usepackage{dcolumn}
\usepackage{bm}
\usepackage{blindtext}
\usepackage{xcolor}
\usepackage{subcaption}
\usepackage{enumerate}
\usepackage{soul}
\def\HLTP{\hat{\mathcal{H}}^{\mathrm{LTP}}}
\def\ansp#1{\hat{a}_{#1}}
\def\crsp#1{\hat{a}_{#1}^{\dag}}
\def\anlo#1{\hat{b}_{#1}}
\def\crlo#1{\hat{b}_{#1}^{\dag}}

\newcommand{\freqn}[2]{\omega_{#2}^{\mathrm{#1}}}

\newcommand{\Eq}[1]{Eq.~\ref{#1}}
\begin{document}

\setstcolor{black}

\preprint{APS/123-QED}

\title{Electrical Generation of Surface Phonon Polaritons}

\author{Christopher R. Gubbin}
\author{Simone De Liberato}%
 \email{s.de-liberato@soton.ac.uk}
\affiliation{School of Physics and Astronomy, University of Southampton, Southampton, SO17 1BJ, United Kingdom}

\date{\today}

\begin{abstract}
Efficient electrical generation of mid-infrared light is challenging because of the dearth of materials with natural dipole-active electronic transitions in this spectral region. One approach to solve this problem is through quantum-engineering of the electron dispersion to create artificial transitions, as in quantum cascade devices. In this work we propose an alternative method to generate mid-infrared light, utilizing the coupling between longitudinal and transverse degrees of freedom due to the nonlocal optical response of nanoscopic polar dielectric crystals. Polar crystals support sub-diffraction photonic modes in the mid-infrared. They also support longitudinal phonons, which couple efficiently with electrical currents through the Fr\"ohlich interaction. As we have shown in previous theoretical and experimental works, these two degrees of freedom can hybridize forming longitudinal-transverse polaritons. Here we theoretically demonstrate that longitudinal-transverse polaritons can be efficiently generated by electrical currents, leading to resonant narrowband photonic emission. 
This approach can therefore be utilised to electrically generate far-field mid-infrared photons in the absence of dipole-active electronic transitions, potentially underpinning a novel generation of mid-infrared optoelectronic devices. 
\end{abstract}

\maketitle

\section*{Introduction}
Surface phonon polaritons (SPhPs) are hybrid light-matter excitations formed from the coupling of free-photons to optical phonons in a polar dielectric crystal \cite{Hillenbrand2002,Caldwell2015a}. They are highly promising for mid-infrared nanophotonics because, like plasmons in the visible, they are morphologically dependent sub-diffraction excitations. This means they are largely tuneable \cite{Sumikura2019, Spann2016, Ellis2016,Gubbin2017,Dubrovkin2020, Wang2013} and have great potential in any application which benefits from a strong electric field such as sensing \cite{Berte2018}, nonlinear optics \cite{Gubbin2017b,Razdolski2018,Kitade2021} or near field imaging \cite{Taubner2006, Kiessling2019}. Novel properties of hyperbolic SPhPs \cite{Caldwell2014, Dai2014, Li2015, Woessner2015, Basov2016, Ma2018, Autore2018} allow for great flexibility in the design of mid-infrared SPhP based devices compared to systems operating in the visible \cite{He2022}. Moreover SPhP resonances are derived from the lattice phonons, and do not require a high electron density allowing them to have significantly narrower linewidths than mid-infrared plasmonic alternatives \cite{Taliercio2019}.\\
In the field of optoelectronics \cite{Gubbin2022} a key proposed application of SPhPs is in mid-infrared thermal emission. Through Kirchhoff's law their intrinsically narrow linewidths allow for the design of narrowband thermal emitters with directional and spectrally tuneable emission \cite{Greffet2002,Schuller2008,Arnold2012,Lu2021}. Despite these impressive results thermal emitters run on spontaneous emission from the thermally oscillating charges of the polar lattice and cannot achieve substantial temporal coherence or lasing. Moreover, the only way to enhance emission is by increasing the device temperature which leads to large short wavelength emission and diminished efficiency in the target spectral range.\\ 
Resonant electrical injection would solve these problems but materials do not typically have strong interband transitions in the mid-infrared and optoelectronic devices operating there have to rely on artificial electronic transitions, created through bandgap engineering \cite{Faist1994, Faist1997, Ohtani2019}. \\
Recent works in surface phonon polariton physics have suggested a solution to this problem. {

While longitudinal and transverse excitations are orthogonal and thus non-interacting in bulk linear polar crystals, in systems with broken translational symmetry they are able to hybridise through common boundary conditions on the electrical and mechanical fields. This effect was recently demonstrated in a study of hybridization between transverse SPhPs localised in 4H-SiC nanopillar resonators and zone-folded longitudinal optical (LO) phonons \cite{Gubbin2019}. The resulting excitations were termed longitudinal-transverse polaritons (LTPs): a reference to the hybrid nature of their electric field and to the resulting anti-crossing in the modal dispersion around the LO phonon frequency. {\color{black} Our theory is a version with retardation of Ridley's one \cite{Ridley1992,Ridley1993,Ridley2009}, in which the role of interface polaritons is played by SPhPs and the role of hybridons by LTPs. The inclusion of retardation in our theory} allows us to describe nanostructured systems in which light can be outcoupled by arrays of defects \cite{Gubbin2016}.
Importantly, the longitudinal polarization field of the LO component can couple directly to charge currents transiting through the lattice. The electrical generation of localised phonons in polar nanolayers has been studied previously in the context of both optical \cite{Constantinou1990, Riddoch1983, Ridley2009, Ridley2017} and acoustic phonons \cite{Wendler1988}. It is known to lead to measurable effects such as velocity saturation in semiconductor devices \cite{Ridley2004, Khurgin2016}. It has also previously been predicted that coherent generation of localised LO phonons could be possible in GaN quantum wells \cite{Komirenko2000, Komirenko2001}.\\ 
In this Paper we explore the idea that the unique \emph{hybrid} nature of LTPs can allow them to be utilised as interconnects between electronic and photonic degrees of freedom in optoelectronic devices. We develop a transparent theory of LTP-driven Fr{\"o}hlich scattering in a planar device, calculating the efficiency of SPhP emission relative to the thermal case. We demonstrate that electrically excited emission is possible, suggesting that LTP interconnects could be widely useful in optoelectronic devices operating across the mid-infrared.}

\section{Overview}
\label{sec:overview}
\begin{figure}
	\centering
	\includegraphics[width=\columnwidth]{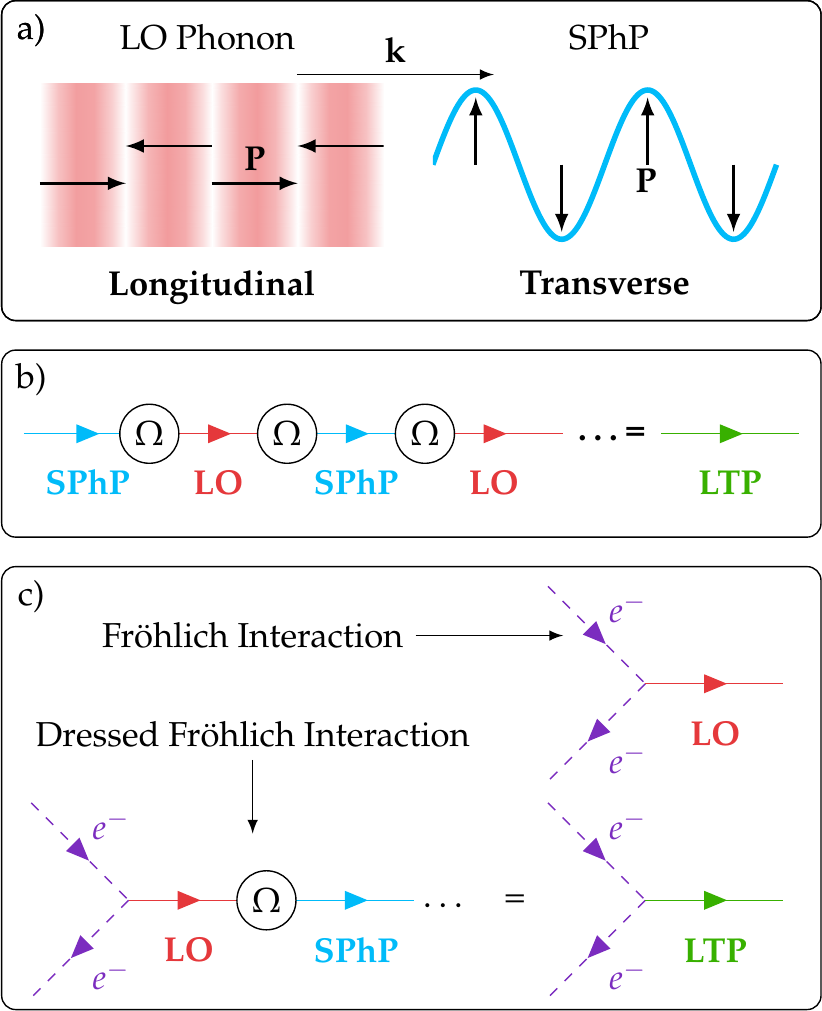}	
	\caption{a) Comparison of the polarization field $\mathbf{P}$ for a longitudinal optical (LO) phonon and a transverse surface phonon polariton (SPhP). In the former the polarization field is parallel to the wavevector $\mathbf{k}$, while in the latter it is orthogonal to it. b) Illustration of strong coupling between LO and SPhP modes. Energy is passed coherently at frequency $\Omega$ between the two degrees of freedom, leading to the formation of a hybrid LTP mode. c) Diagrammatic sketch of how the coupling between LO phonons and SPhPs can dress the Fr{\"o}hlich interaction between electrons and LO phonons, leading to a resonant interaction between electrons and LTPs.}
	\label{fig:1}
\end{figure}
Polar crystals support short-wavelength LO phonons. These are dispersive, meaning their frequency is a function of wavevector $\omega_{\mathrm{L}} \left(\mathbf{k}\right)$. { Longitudinal phonons are characterized by a curl-free  polarization field, $\nabla\times\mathbf{P}=0$, which in bulk implies $\mathbf{P}$ is parallel to the propagation vector $\mathbf{k}$, so $\mathbf{k} \times \mathbf{P} = 0$, as illustrated in Fig.~\ref{fig:1}a. Optical phonons at the surface of a polar crystal can also couple to photons, forming sub-diffraction optical modes termed surface phonon polaritons (SPhPs), which exist in the Reststrahlen region between the lattice longitudinal and transverse optical phonon frequencies. In this spectral window a polar crystal is characterized by a negative dielectric function and responds like a metal to external electromagnetic fields. In local optical theories the polarization field of SPhPs is {\it almost everywhere} divergence free, $\nabla\cdot\mathbf{P}=0$, excepting the interfaces of the polar material. In this Paper we refer to SPhPs as transverse excitations, signifying that their polarization field and complex wavevectors are orthogonal in each medium ($\mathbf{k} \cdot \mathbf{P} = 0$), as schematically shown in Fig.~\ref{fig:1}a.
SPhPs are purely transverse because local optics parameterizes the transverse dielectric function of the polar lattice with only the zone-center $\mathbf{k} = 0$ phonon frequencies, neglecting dispersion of, and energy transport in, the optical phonon branches.} As we demonstrated in a recent series of publications, which allowed us to quantitatively reproduce previously unexplained experimental data \cite{Gubbin2020, Gubbin2020b, Gubbin2020c, Gubbin2022b, Gubbin2022c}, accounting for the optical phonon dispersion leads to a \emph{nonlocal} theory of polar optics analogous to that in plasmonic systems \cite{Mortensen2014, GarciadeAbajo2008, David2011, Raza2011, Ciraci2012} in which LO phonons and SPhPs interact through shared electro-mechanical boundary conditions. In large resonators the effect is an increased non-radiative damping \cite{Gubbin2020} as propagative phonons leach energy from the SPhP, but in nanoscale systems the interaction can become coherent. In this regime energy passes multiple times between LO phonons and SPhPs before dissipating so the two modes can no longer be considered distinct. Instead they are best understood as a hybrid excitation, termed longitudinal-transverse polariton (LTP), illustrated in Fig.~\ref{fig:1}b, whose electric field is a linear mixture of each component. The crossover point between these regimes occurs where the thickness of the polar medium approaches the longitudinal phonon propagation length in the material and a discrete LO phonon spectrum emerges. At this limit an appreciable fraction of LO phonons excited at the edge of the polar crystal are able to transit the layer before decaying non-radiatively, allowing their energy to recycle into SPhPs.\\
\begin{figure}
	\centering
	\includegraphics[width=0.75\columnwidth]{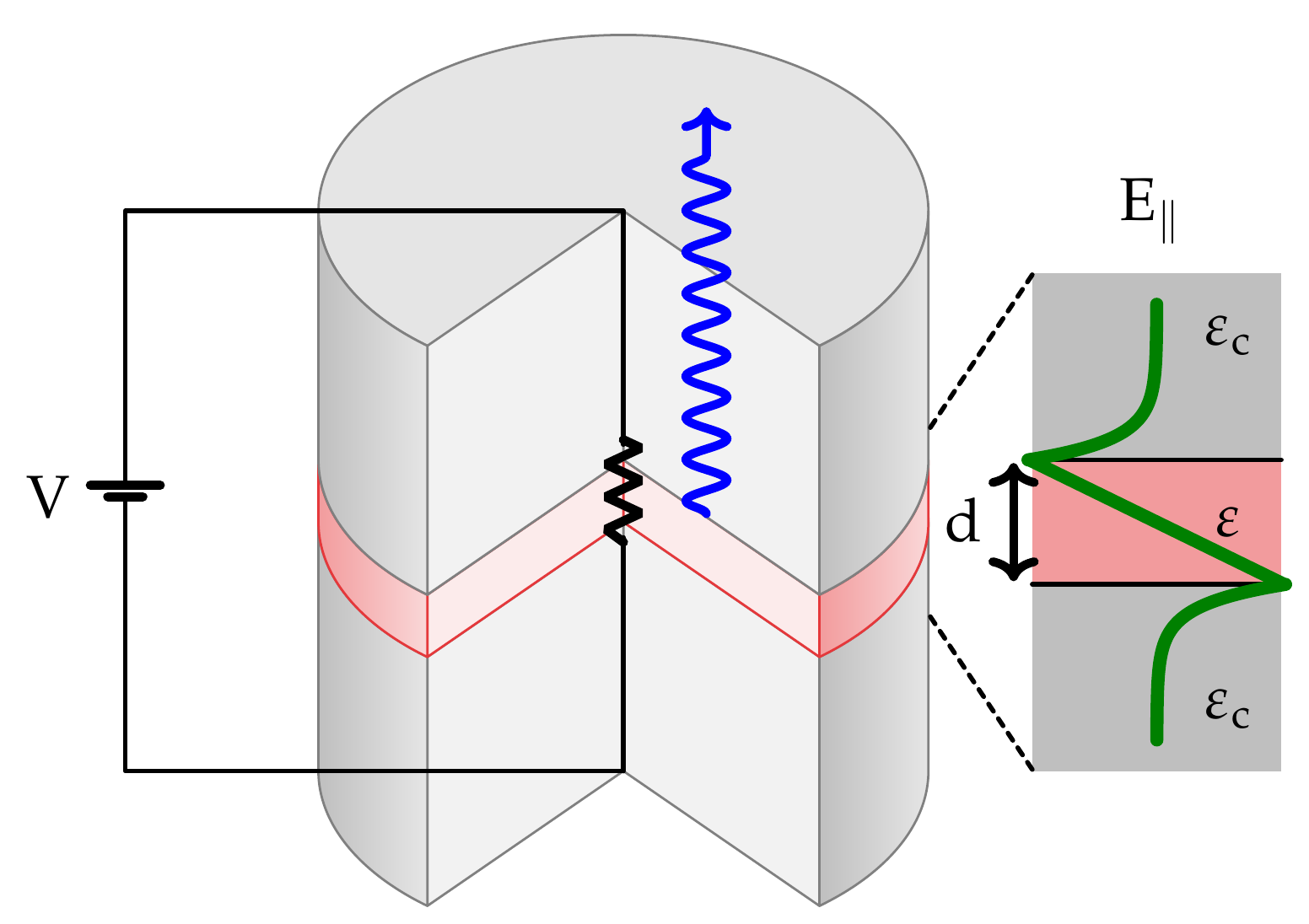}	
	\caption{Sketch of the device studied in this Paper. SPhPs supported by a polar layer of thickness $d$ and dielectric function $\epsilon$ in a cladding semiconductor with dielectric constant $\epsilon_{\mathrm{c}}$ are excited by application of a voltage across the film. The SPhPs are then emitted at a rate proportional to their photonic fraction. }
	\label{fig:2}
\end{figure}\\
 Longitudinal phonons can be excited by electrons passing through a polar crystal. Charged electrons perturb the ions of the crystal lattice from their equilibrium positions. This emission of LO phonons, named after Fr{\"o}hlich, is one of the leading dissipation channels for Ohmic losses and is illustrated by the upper diagram in Fig.~\ref{fig:1}c. In a system supporting LTPs the emitted LO phonons can convert energy into SPhPs, and following the scheme in Fig.~\ref{fig:1}b this leads to a \emph{dressed} Fr{\"o}hlich interaction in which electrons resonantly emit LTPs, illustrated in the lower diagram Fig.~\ref{fig:1}c. 
As SPhPs can couple to the external environment through nanophotonic elements, emitting radiation in the far-field, this mechanism is able to act as an \emph{interconnect} between electronic and photonic sub-systems, allowing injected electrical currents to radiate photons in the far-field.\\
\begin{figure*}
	\centering
	\includegraphics[width=\textwidth]{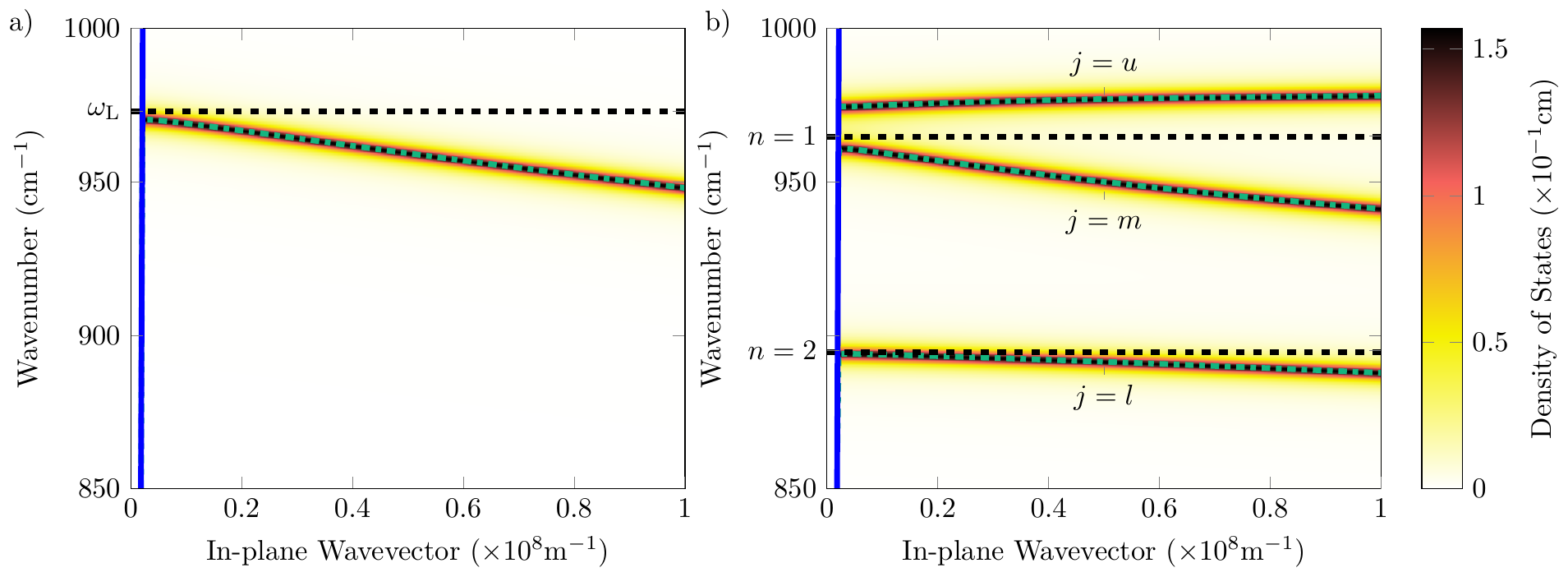}	
	\caption{Density of states and modal dispersions for a 3C-SiC film of thickness 2nm characterised by \Eq{eq:Lorentz}, sandwiched in a Si cladding ($\epsilon_{\mathrm{c}} = 11.71$).  In panel a), representing the prediction of a local theory, the black dashed line shows the bulk LO phonon frequency $\omega_{\mathrm{L}}$ and the dot-dashed green line the local SPhP dispersion $\omega_{\mathbf{q}}^{\mathrm{SP}}$(Appendix~\ref{app:spectral}). In panel b), representing the prediction of a nonlocal theory, the black dashed lines shows the localised LO phonon frequencies $\omega_{\mathbf{q},n}^{\mathrm{L}}$ and the dot-dashed green lines the LTP frequencies $\omega_{\mathbf{q},j}^{\mathrm{LTP}}$. In both panels blue solid lines mark edge of the light-cone in the cladding.}
	\label{fig:3}
\end{figure*}
In this Paper we study the simple example of an LTP interconnect consisting of a thin polar film of thickness $d$ embedded in a bulk material with dielectric constant $\epsilon_{\mathrm{c}}$, sketched in Fig.~\ref{fig:2} \cite{Gubbin2020c}. For simplicity we calculate the modes of our device considering infinite lateral extent.  We consider the polar layer to be cubic silicon carbide (3C-SiC), characterized {in the local limit} by the Lorentzian dielectric function
\begin{equation}
	\epsilon\left(\omega\right) = \epsilon_{\infty} \frac{\omega_{\mathrm{L}}^2 - \omega\left(\omega + i \gamma\right)}{\omega_{\mathrm{T}}^2 - \omega \left(\omega + i \gamma\right)}, \label{eq:Lorentz}
\end{equation}
where $\omega_{\mathrm{L}} = 973\mathrm{cm}^{-1}$ is the zone-center LO phonon frequency, $\omega_{\mathrm{T}} = 796\mathrm{cm}^{-1}$ is the zone-center TO phonon frequency, $\gamma = 4\mathrm{cm}^{-1}$ is the non-radiative damping rate, and $\epsilon_{\infty} = 6.52$ is the high-frequency dielectric constant \cite{Moore1995}. The cladding is taken as silicon (Si), characterized by the dielectric constant $\epsilon_{\mathrm{c}} = 11.71$.\\
The polar layer supports a SPhP mode at each interface which are uncoupled in a thick film. When $d$ approaches the skin-depth of the SPhP they begin to hybridize forming two polariton branches, one with frequency below the uncoupled mode frequency, and one with frequency above \cite{maierbook}. In this work we concern ourselves with the high-frequency SPhP whose frequency is close to the zone-center longitudinal phonon frequency \cite{Passler2018}. In Appendix~\ref{app:spectral} the complex frequency of this mode is derived in the local-response-approximation, where the dispersion of the underlying phonon modes of the lattice is disregarded. In Fig.~\ref{fig:3}a we plot the local density of states which describes the SPhP dispersion
\begin{equation}
	\rho_{\mathbf{q}}^{\text{SP}} \left(\omega\right) = \frac{1}{\pi} \frac{\Im \left\{ \omega_{\mathbf{q}}^{\mathrm{SP}} \right\}}{\left(\omega - \Re \left\{ \omega_{\mathbf{q}}^{\mathrm{SP}} \right\}\right)^2 + \Im \left\{ \omega_{\mathbf{q}}^{\mathrm{SP}} \right\}^2}, \label{eq:locdos}
\end{equation}
in which $\omega_{\mathbf{q}}^{\mathrm{SP}}$ is the SPhP frequency as a function of in-plane wavevector $\mathbf{q}$ for a layer thickness $d=2$nm. At small wavevectors the mode is in the epsilon-near-zero (ENZ) regime \cite{Campione2015, Khurgin2021} where the SPhP phase shift over the film thickness is negligible, and the frequency lies near the zone-center LO phonon frequency $\omega_{\mathrm{L}}$, indicated by the black dashed line in Fig.~\ref{fig:3}a. At large wavevectors this is no longer the case, the SPhPs on each interface of the polar layer begin to decouple and the mode red shifts toward the asymptotic SPhP frequency of the bilayer which it reaches when $\mathbf{q}$ is sufficiently large that the SPhPs on each interface no longer overlap. The real modal frequency $\Re \left\{ \omega_{\mathbf{q}}^{\mathrm{SP}} \right\}$ is indicated by the green dash-dotted line. Note that for simplicity in Fig.~\ref{fig:3} and for the remainder of this Paper we ignore the lower energy SPhP mode. This excitation, whose small-wavevector dispersion begins at $\omega_{\mathrm{T}}$ approaches the same asymptotic frequency from the low-wavenumber side in the large wavevector limit.\\
A thin film whose thickness approaches the LO phonon propagation length also acts as a Fabry-P\'erot resonator for LO phonons, leading to the appearance of a discrete spectrum of longitudinal modes with quantized modal frequencies
\begin{equation}
	\omega_{\mathbf{q}, n}^{\mathrm{L}} = \sqrt{\omega_{\mathrm{L}}^2 - \beta_{\mathrm{L}}^2 \left(\lvert \mathbf{q} \rvert^2 + \xi_n^2\right)},	\label{eq:LOFreq}
\end{equation} 
in which $n\in\mathbb{N}$ labels the discrete phonon branch, $\beta_{\mathrm{L}}  = 15.3 \times 10^{5} \mathrm{cm \,s}^{-1}$ is the LO phonon velocity in the low-wavevector quadratic regime \cite{Karch1994, Gubbin2020} and the quantized out-of-plane wavevector of the $n$th localised phonon mode is given by
\begin{equation}
    \xi_n = \frac{(2 n - 1) \pi}{d}. \label{eq:LOwv}
\end{equation}
The factor $2 n - 1$ here ensures only LO phonons with equal parity to the high-frequency SPhP branch are considered, as in the symmetric structure under consideration only these modes are coupled. The remaining phonon modes couple to the low-frequency SPhP and are not considered in this Paper.\\
{ In the full nonlocal problem the transverse and longitudinal electromagnetic fields in the waveguide must be calculated simultaneously. As the standard Maxwell boundary conditions on the transverse components of the electric and magnetic fields leave an under-specified problem this requires the application of additional boundary conditions. We have previously derived the appropriate additional conditions on the mechanical fields in the polar layer considering continuity of the nonlocal Poynting vector, which accounts for energy carried in both the electromagnetic and mechanical fields, across each interface \cite{Gubbin2020}. This imposes continuity of the lattice displacement field $\mathbf{X}$ and the normal component of the lattice stress tensor $\bar{\boldsymbol{\tau}}$. Note that in this work we ignore TO phonon modes in the polar material, an approximation we previously demonstrated was valid in SiC because of the large LO-TO splitting \cite{Gubbin2020c}.  In this limit the problem is over-specified and it is not possible to utilize all the additional boundary conditions, we choose to apply the Pekar-Ridley condition \cite{Pekar1959} $\mathbf{X} = 0$ at the interfaces of the polar film. Under these approximations it} is possible to derive the analytic expression for the LO-SPhP coupling frequency \cite{Gubbin2022b}
\begin{equation}
	\lvert \Omega_{\mathbf{q}, n} \rvert^2 =  \frac{2 \beta_{\mathrm{L}}^2}{d^2} \frac{\omega_{\mathrm{L}}^2 - \omega_{\mathbf{q}}^{\mathrm{SP}\;2}}{\omega_{\mathbf{q}}^{\mathrm{SP}} \omega_{\mathbf{q}, n}^{\mathrm{L}}}.\label{eq:couplinga}
\end{equation}
The LO-SPhP coupling is mediated by interactions between the two excitations at the surface of the polar layer, meaning it is highly sensitive to the time the LO phonon takes to propagate across the film. For this reason it is diminished in thicker films, and for modes close to the zone-center LO frequency where the phonon group velocity is small. \\
For a $2$nm 3C-SiC film we observe two localised LO modes in the spectral window under study, illustrated by the the $n=1$ and $n=2$ dashed black lines in Fig.~\ref{fig:3}b. The LTP modes are linear superpositions of the underlying SPhP and LO phonon modes coupled at frequencies given by \Eq{eq:couplinga}, whose eigenfrequencies can be derived explicitly by Hopfield diagonalization \cite{Hopfield1958} as discussed in Appendix~\ref{app:spectranl}. The complex LTP frequencies are given by $\omega_{\mathbf{q}, j}^{\mathrm{LTP}}$ where $j$ labels the discrete polariton branches.
The total number of LTP branches is equal to the number of bare modes: one greater than the number of localised LO phonons. We find three modes: an almost dispersionless upper polariton $\omega > \omega_{\mathrm{L}}$ ($j=u$), a middle branch ($j=m$) sandwiched between the localised LO frequencies, and a lower polariton branch ($j=l$) which tends to the the ENZ modal frequency at large in-plane wavevectors.
The multimode analogue of \Eq{eq:locdos} is plotted in the colormap Fig.~\ref{fig:3}b, while real modal frequencies $\Re\left\{\omega_{\mathbf{q}, j}^{\mathrm{LTP}}\right\}$ are shown with dash-dotted green lines and marked by the relative value of $j$. \\
The nonlocal physics which leads to LTP formation can be understood considering the dispersion of the LO phonon branch in the lattice. Away from the zero-wavevector $\Gamma$ point the band red-shifts. At each frequency the LO phonon mode has a unique wavelength, calculable from the dispersion relation, and when an integer multiple of this wavelength fits within the polar film a Fabry-P{\'e}rot mode forms. These discrete modes couple to the SPhP, leading to the altered dispersion in Fig.~\ref{fig:3}b. The wavevectors of the three localised modes participating in the coupling satisfy the resonance condition in \Eq{eq:LOwv}.\\

The process of LTP-driven electrically excited emission sketched in Fig.~\ref{fig:1} has many moving parts which must be taken into account to correctly model the conversion of electrical to photonic energy. Electrons are driven by externally applied voltages and can radiate LTPs which can then either decay radiatively emitting photons into the far-field or non-radiatively. To arrive at a fundamental and intelligible understanding of the potential of LTP interconnects for the design of mid-infrared optoelectronic devices we keep the full microscopic complexity of the LTP physics and make important simplifications to the description of the couplings to the external environment:
 
\begin{enumerate}[I]
\item {  We do not utilize a self-consistent out-of-equilibrium many-body theory of the electrical transport [59]. Instead we consider an electron gas at thermal equilibrium with a well defined temperature $T_e$ described by a Maxwell-Boltzmann distribution
\begin{equation} \tag{6}
f_{\vec{k}}\left(T_e\right) = \frac{8\pi^3}{\mathrm{V}} \frac{ e^{- \varepsilon_{\vec{k}} / k_{\mathrm{B}} T_e}}{N_\mathrm{c} \left(T_e\right)},
\end{equation}
where $N_\mathrm{c}\left(T_e\right) = 4 \eta \left[\frac{m^* k_{\mathrm{B}} T_e}{2 \pi \hbar^2}\right]^{3/2}$ is the density of states in the conduction band, $\eta$ is the degeneracy factor in the conduction band, $\mathrm{V}$ is the electronic quantisation volume, $\vec{k}$ is the three-dimension electron wavevector and $\varepsilon_{\vec{k}}$ is the electron energy. In Appendix C we consider the balance between electron energy and drift velocity in the thermalized electron gas, demonstrating that both the drift velocity and applied electric field between the device contacts can be uniquely determined using only the temperature $T_e$ and carrier density $n_e$ of the gas. In all calculations presented in this Paper the electron density is taken to be $n_e = 10^{18} \; \mathrm{cm}^3$. This density is chosen low enough not to significantly perturb the SPhP resonances [60] and to allow us to consider the gas to be non-degenerate, removing the need to consider Pauli blocking in the outgoing electronic states (see Appendix~C).}

\item LTPs in the system are below-light-line excitations. An outcoupling mechanism must be introduced in order to allow them to radiate in the far-field. We will not evaluate a specific extraction grating or prism geometry, as here our aim is to provide a general evaluation of the efficiency of the underlying emission mechanism. A good estimate of the level at which an LTP is able to couple to the far-field can be obtained by considering how photon-like the mode is. The effective photon population is given by the LTP population times its photonic fraction, equal to the ratio of the group velocity to the speed-of-light in vacuum
\begin{equation}
    \xi_{\mathbf{q}, j} = \frac{1}{c} \frac{\mathrm{d} \omega_{\mathbf{q}, j}^{\text{LTP}}}{\mathrm{d} q}. \label{eq:phohop}
\end{equation}
In this model polaritons with vanishing group velocity, in the flat-band region of the LTP dispersion, have a vanishing coefficient $\xi_{\mathbf{q}, j}$ as they are essentially just localised LO phonon modes decoupled from the SPhP. Those in regions near anti-crossings achieve instead a finite $\xi_{\mathbf{q},j}$ from their SPhP component and for suitably small values of the wavevector they can be extracted using standard and relatively efficient gratings or prisms. 

\end{enumerate}

We are now able to formulate a rate equation for the LTP populations $N_{\mathbf{q}, j}$ in the presence of both thermal and electrical pumping
\begin{multline}
\dot{N}_{\mathbf{q}, j}=\gamma^{\text{LTP}}_{\mathbf{q}, j}\left[ N^{\text{TH}}_{\mathbf{q}, j}(T_l)-N_{\mathbf{q}, j}\right] 
+\Gamma_{\mathbf{q}, j}^{\text{LTP}}(N_{\mathbf{q}, j},T_e), \label{eq:bose}
\end{multline}
where $\gamma^{\text{LTP}}_{\mathbf{q}, j}$, defined in Appendix~\ref{app:spectranl} is the branch- and momentum-resolved LTP damping and 
\begin{align}
\label{eq:Gamma}
\Gamma_{\mathbf{q}, j}^{\text{LTP}}(N_{\mathbf{q}, j},T_e)&=  \Gamma_{\mathbf{q}, j}^{+}(T_e)(1+N_{\mathbf{q}, j})-\Gamma_{\mathbf{q}, j}^{-}(T_e)N_{\mathbf{q}, j},
\end{align}
is the term describing LTP emission and reabsorption by the electron gas, depending on both the electronic temperature $T_e$ and the LTP population. The emission ($+$) and absorption ($-$) coefficients are defined in Appendix~\ref{app:Thermal}. 
When the lattice is at equilibrium with the electron gas the LTP population remains stationary, implying that $\Gamma_{\mathbf{q}, j}^{\text{LTP}}(N_{\mathbf{q}, j}^{\text{TH}}(T),T)=0$. 
We do not mark explicitly the dependence upon the electron density $n_e$, but both $\Gamma_{\mathbf{q}, j}^{+}(T_e)$ and $\Gamma_{\mathbf{q}, j}^{-}(T_e)$ are linearly proportional to it. 
In \Eq{eq:bose} we also introduced a finite lattice temperature $T_l$ to account for thermal LTP occupation, with equilibrium thermal distribution 
\begin{equation}
   N^{\text{TH}}_{\mathbf{q}, j}(T) = \left[ \exp\left(\frac{\hbar\omega_{\mathbf{q}, j}^{\textrm{LTP}} }{k_{\mathrm{B}} T}\right) - 1 \right]^{-1}. \label{eq:anin}
\end{equation}
The population in the steady-state $\dot{N}_{\mathbf{q}, j} = 0$ is therefore given by
\begin{align}
 N^{\text{ss}}_{\mathbf{q}, j}(T_e,T_l)&=\frac{\gamma^{\text{LTP}}_{\mathbf{q}, j} N^{\mathrm{TH}}_{\mathbf{q}, j} \left(T_l\right) + \Gamma_{\mathbf{q}, j}^{\mathrm{+}}(T_e)}{\gamma^{\text{LTP}}_{\mathbf{q}, j} + \Gamma_{\mathbf{q}, j}^{\mathrm{-}}(T_e) - \Gamma_{\mathbf{q}, j}^{\mathrm{+}}(T_e)}. \label{eq:ss}
\end{align}
We also introduce a related quantity, the spectral steady-state population, which allows to take into account the finite linewidth of the LTP resonances. This relates to the branch-resolved population through 
\begin{align}
 N^{\text{ss}}_{\mathbf{q}, j} \left(\omega,T_e,T_l\right)&= N^{\text{ss}}_{\mathbf{q}, j} \left(T_e,T_l\right) \rho_{\mathbf{q}, j}^{\text{LTP}} \left(\omega\right), \label{eq:ssrho}
\end{align}
in which $\rho_{\mathbf{q}, j}^{\text{LTP}}\left(\omega\right)$ is the density of LTP states at wavevector $\mathbf{q}$ in branch $j$ at frequency $\omega$. This density is modeled as a Lorentzian and is defined in Appendix A2.

Note that our rate-equation does not account for Joule heating of the phonon reservoir. Although in a realistic device the temperature of the electronic sub-system $T_e$ and of the lattice $T_l$ would evolve toward an equilibrium over time, describing this process is beyond the scope of this Paper as the dynamics would be device- and operation-dependent. 

\section{Results}
\begin{figure*}
    \begin{subfigure}[b]{0.8\linewidth}
        \includegraphics[width=\textwidth]{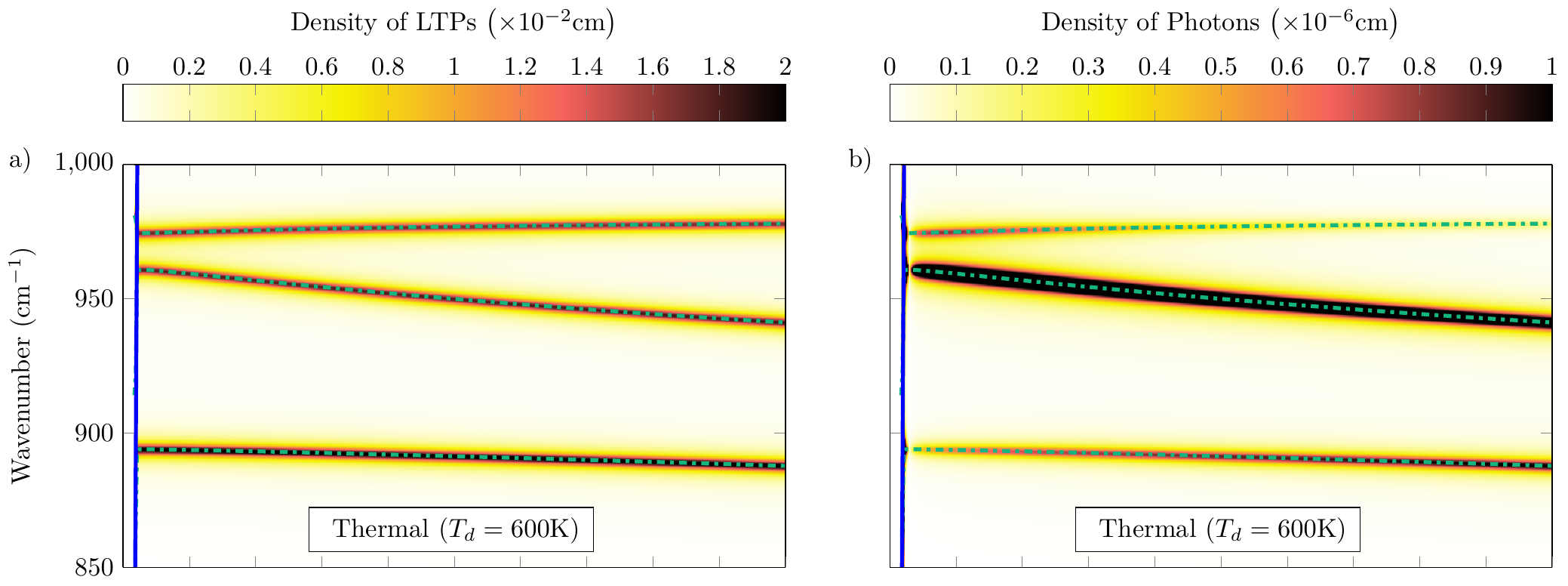}
    \end{subfigure}
    \begin{subfigure}[b]{0.8\linewidth}
        \includegraphics[width=\textwidth]{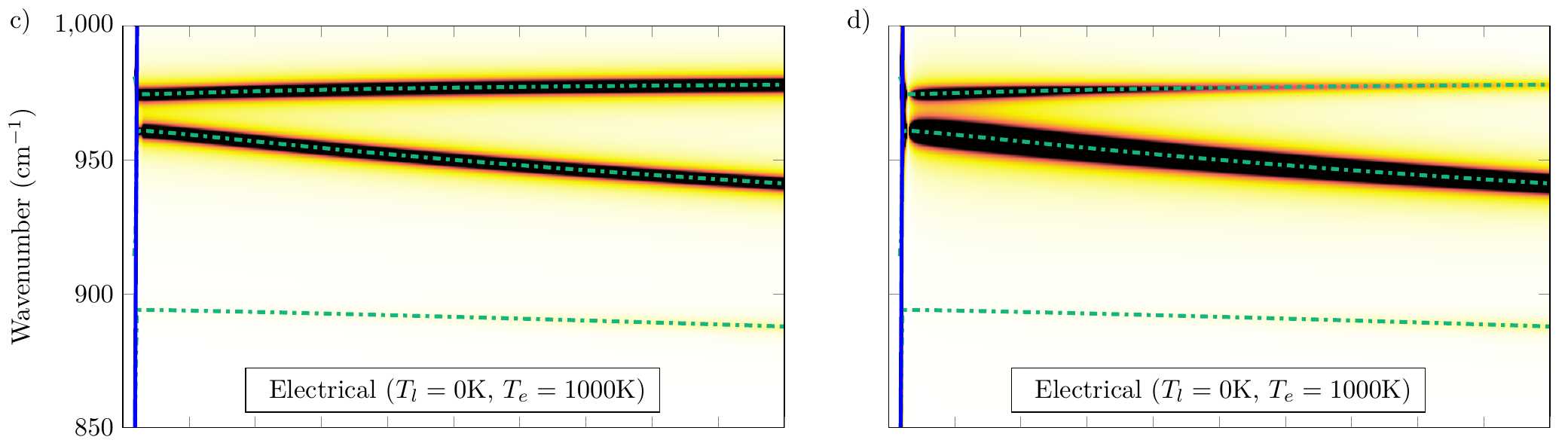}
    \end{subfigure}
    \begin{subfigure}[b]{0.8\linewidth}
        \includegraphics[width=\textwidth]{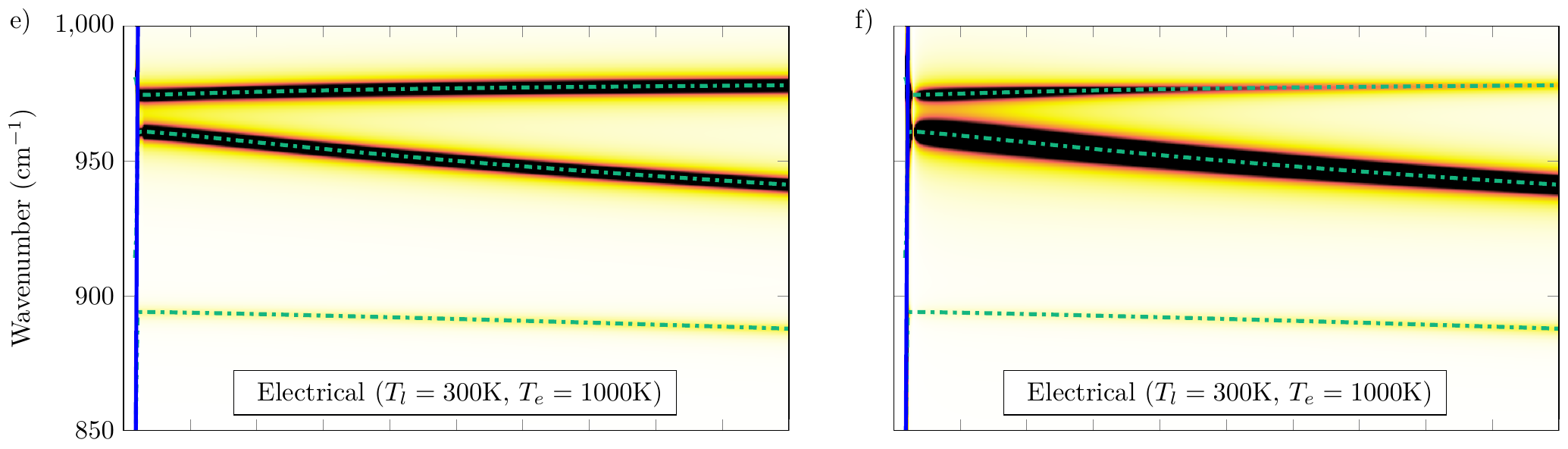}
    \end{subfigure}
    \begin{subfigure}[b]{0.8\linewidth}
        \includegraphics[width=\textwidth]{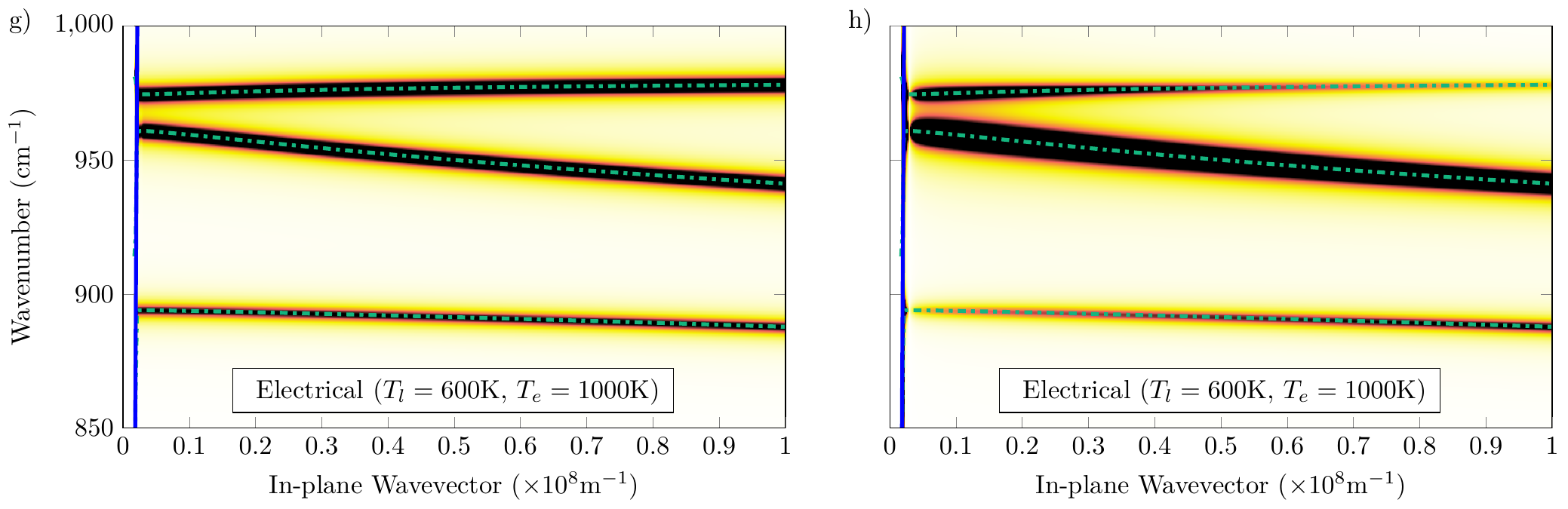}
    \end{subfigure}
   	\caption{{\color{black}Comparison of LTP and photonic steady-state populations generated by thermal and electrical processes. In the first column we plot the LTP density, while in the second column we plot the photonic density, obtained by multiplying the LTP density by the photon fraction $\xi_{\mathbf{q},j}$ from \Eq{eq:phohop}. The first row (a,b) shows data obtained considering purely thermal emission, obtained setting both the lattice and the electronic temperatures to the same equilibrium device temperature $T_l = T_e = 600$K. 
    The other three rows show instead data obtained considering electronic temperature $T_e=1000$K and lattice temperatures equal to $T_l=0$K (c,d), $T_l=300$K (e,f), and $T_l=600$K (g,h). 
   Other parameters as in Fig. \ref{fig:3}. In all panels the blue-solid line indicates the light-cone edge in the cladding, and the dot-dashed green lines indicate the real LTP frequencies. Note that the color in the figure has been saturated to allow the thermal and the much larger electrical populations to be plotted on the same color scale.}}
   	\label{fig:4}
\end{figure*}
LTP-driven electrically excited emission  is the result of two chained processes: incoupling from electronic current to LTP modes and subsequently outcoupling to the far-field. The former depends on the density of initial and final states within the electron gas, and is enhanced at moderately large wavevectors where the polariton is phonon-like but the transferred wavevector is not sufficient to prohibit phase-matching. This differs from a thermal excitation which only varies according to the modal frequency and the corresponding Bose-Einstein population $N^{\mathrm{TH}}_{\mathbf{q},j}(T_l)$ at the temperature of the crystal lattice. \\
We are now in a position to estimate the efficiency of the LPT interconnect, by comparing it with the same device operated as a thermal emitter. To simulate thermal emission we consider the system at the equilibrium device temperature $T_l=T_e=600$K, a temperature typical for SiC-based thermal emitters \cite{Lu2021}.
{\color{black} For the electrically excited emission we consider instead a fixed electronic temperature $T_e = 1000$K, a temperature chosen to have electrons propagating near their saturation velocity (Appendix C), and three different lattice temperatures: 
$T_l=0,300,$ and $600$K. The results corresponding to the first value represent the intensity of purely electrically excited emission, allowing us to estimate its intrinsic efficiency when compared to the purely thermal process. The results corresponding to the second and third values provide an estimate for the populations expected in actual devices operating at room temperature and high temperature respectively. 

In the left column of Fig.~\ref{fig:4} we plot the steady-state LTP populations per unit frequency from \Eq{eq:ssrho} in the case of purely thermal (a), purely electrical ($T_l = 0$K) (c), room temperature ($T_l = 300$K) (g), and high temperature ($T_l = 600$K) (h) operation.\\
In the thermal case all regions of the dispersion are similarly populated: because the Reststrahlen region is relatively narrow the thermal occupancy $N^{\mathrm{TH}}_{\mathbf{q},j}(T_l)$ varies slowly over the plotted region. For the purely electrical incoupling (c) LTPs are instead far more likely to be generated in the spectral region near to the zone-center LO phonon frequency $\omega_{\mathrm{L}}$.  
Electrical, finite temperature results (g,h) are intermediate between the previous two: while the dominant emission remains the electrical one near to the LO frequency, additional thermal radiation leads to an increased emission into the lower polariton branch.
In the right column of Fig.~\ref{fig:4} we plot instead the steady state photon population for the same parameters, that is we multiply the results in the left column by the corresponding photon fraction from \Eq{eq:phohop}.} We can see how large photonic components are generated in the smaller wavevector regions, specifically in high-dispersion regions below the polariton stopbands. Note that the photonic populations in this figure go to zero near to the light-line, in the region where the underlying LTP group velocity is near-zero.\\

Regardless of the extraction mechanism used to emit LTPs to the far-field, both electrical and thermal emission in the same mode would be extracted with the same efficiency. Moreover, the wavevector region $q\approx 10^7$, well resolved in Fig.~\ref{fig:4}, only requires standard micrometer-sized gratings for outcoupling. This figure thus provides the main result of this paper, demonstrating that the LTP interconnect we propose can emit more than standard narrow-band SPhP-based thermal emitters \cite{Lu2021}, while greatly reducing the energy dissipated outside the narrow region of interest.\\
 
\begin{figure}
	\centering
	\includegraphics[width=0.95\linewidth]{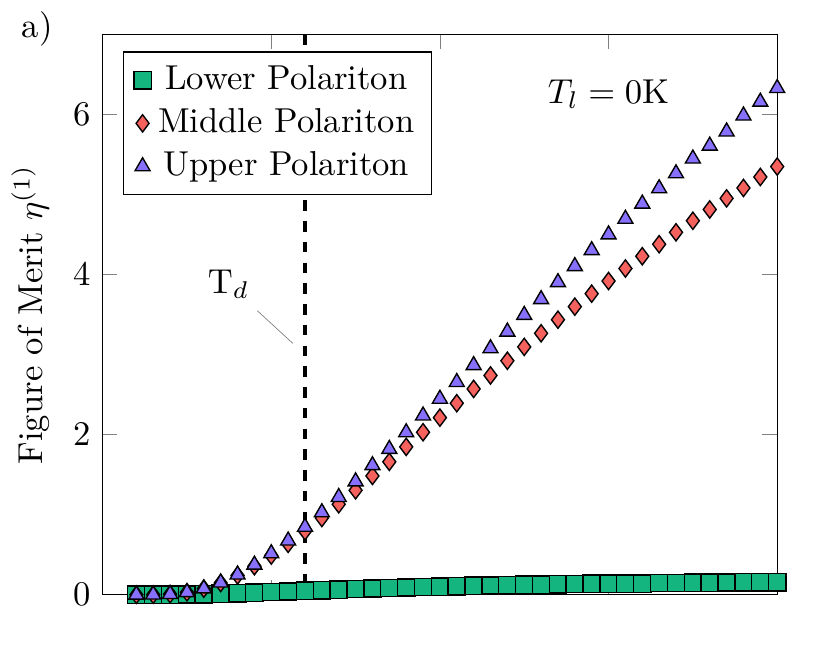}	
	\includegraphics[width=0.95\linewidth]{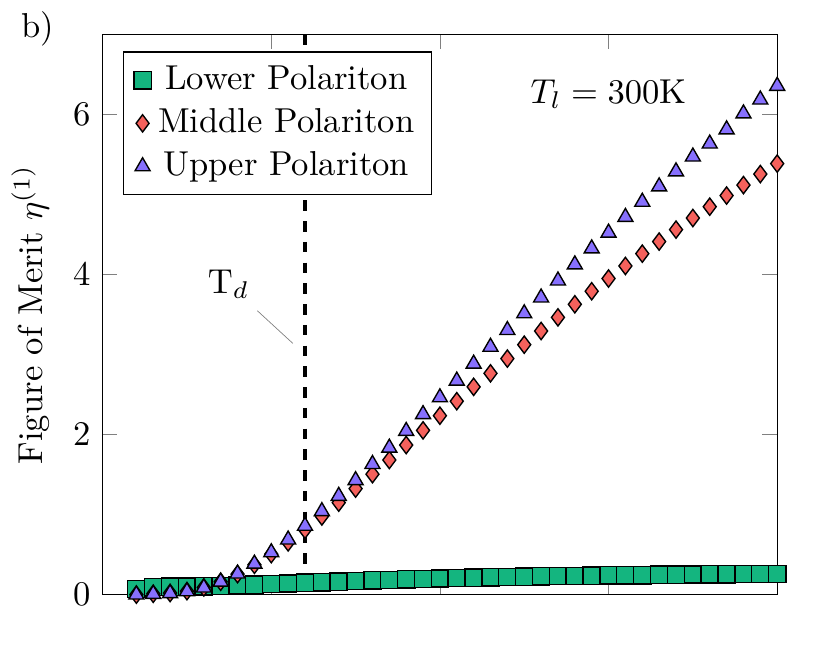}	
	\includegraphics[width=0.95\linewidth]{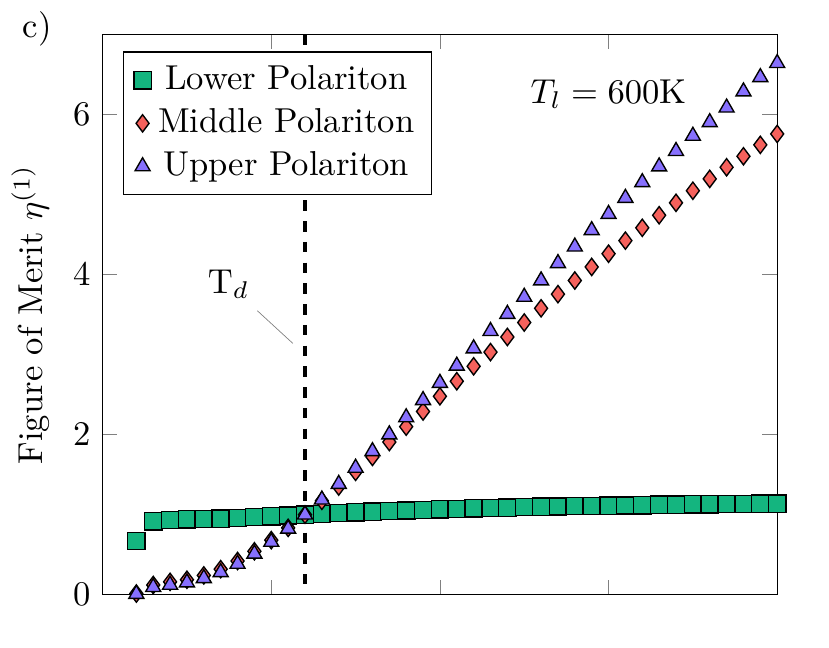}	
	\caption{{\color{black}Figure of merit $\eta_{\mathbf{q}, j}^{(1)}$ from  \Eq{eq:eta1}, as a function of the electronic temperature $T_e$, evaluated at fixed in-plane wavevector $q=10^7 \mathrm{m}^{-1}$, for active region of thickness 2nm, device temperature $T_d = 600$K, and lattice temperatures $T_l = 0$K (a), $T_l = 300$K (b), and $T_l = 600$K (c). For each temperature the figure of merit is plotted for each polariton branch, with reference to the dispersion in Fig.~\ref{fig:3} these correspond to the lower (green squares), middle (red diamonds) and upper (purple triangles) polariton branches.}}
	\label{fig:5}
\end{figure}

In order to proceed with the quantitative comparison between electrical and thermal emission, we consider the figure of merit
{\color{black}
\begin{align}
\label{eq:eta1}
    \eta_{\mathbf{q}, j}^{(1)}\left(T_e, T_l, T_d\right) &= \frac{N_{\mathbf{q}, j}^{\text{ss}}(T_e,T_l)}{N_{\mathbf{q}, j}^{\text{ss}}(T_d,T_d)},  
\end{align}
which is the in-plane momentum $\mathbf{q}$ and branch $j$ resolved ratio between electrically excited emission with electronic temperature $T_e$ and lattice temperature $T_l$, and purely thermal emission at the equilibrum device temperature $T_d$.
In Fig.~\ref{fig:5} we plot the branch resolved values of $\eta_{\mathbf{q}, j}^{(1)}$ as a function of electronic temperature $T_e$, with the in-plane wavevector fixed to $q=10^{7} \mathrm{m}^{-1}$, for a 2nm active region, equilibrium device temperature $T_d=600$K and
$T_l = 0$K (a), $T_l = 300$K (b), and $T_l = 600$K (c). 
As the electronic temperature increases, the magnitude of $\eta_{\mathbf{q}, j}^{(1)}$ increases and the electrically excited emission becomes substantially larger than the thermal one. \\
As already seen in Fig.~\ref{fig:4} the majority of electrical generation happens into the upper-polariton branch (purple triangles) and middle-polariton branch (red diamonds), which are spectrally close to the zone-center LO phonon frequency $\omega_{\mathrm{L}}$, thus 
causing little dissipation outside a specific and narrow frequency region and improving the energy efficiency of the process when compared to the broadband thermal dissipation.
At finite values of $T_l$ (Fig.~\ref{fig:5}b,c) the figure of merit for the middle- and upper-polaritons is slightly enhanced because of the additional thermal emission.
In the lower polariton branch (green squares) the electrically excited emission is instead much smaller due to the strong spectral separation between $\omega_{\mathrm{L}}$ and the polariton frequency \cite{Gubbin2022b}, and at finite $T_l$ it is largely dominated by the thermal component.}

\begin{figure}
	\includegraphics[width=0.95\linewidth]{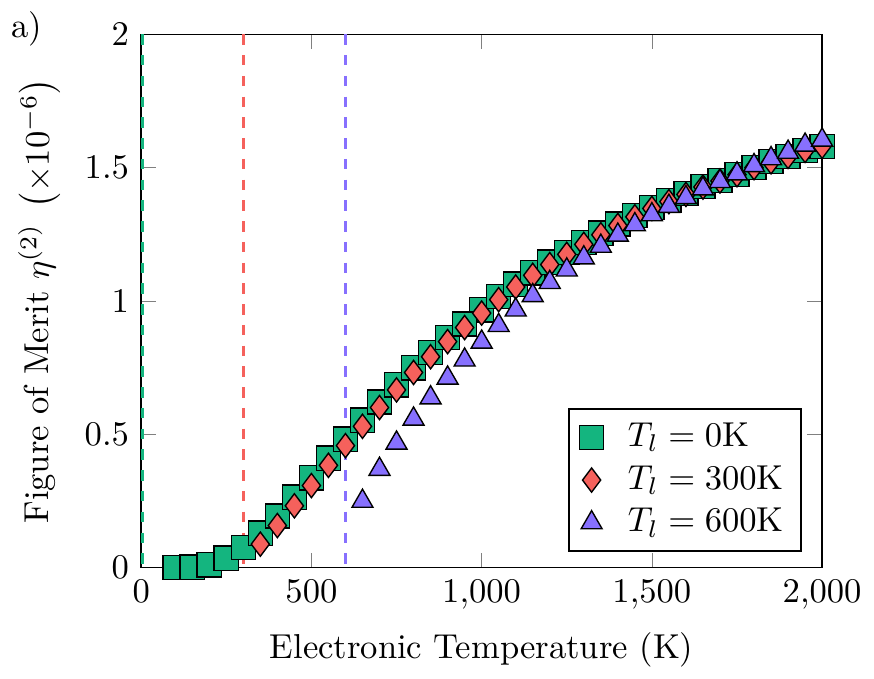}	
	\includegraphics[width=0.9\linewidth]{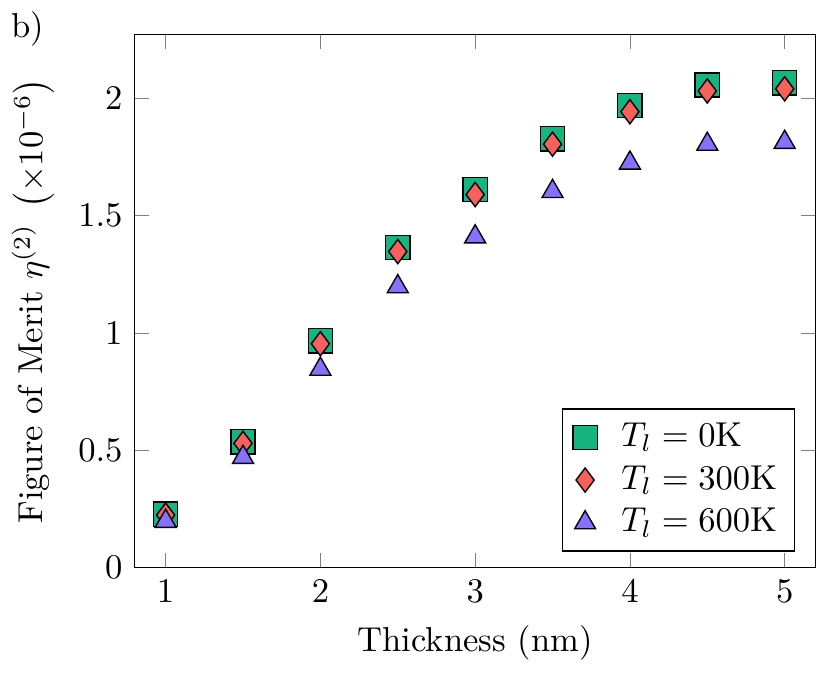}	
\caption{{\color{black} Figure of merit $\eta^{(2)}$. In panel (a) we plot it as a function of the electronic temperature $T_e$, for an active region of thickness 2nm. In panel (b) we plot the same quantity as a function of the thickness for a constant electronic temperature $T_e = 1000$K. Results are plotted for lattice temperature $T_l = 0$K (green squares),  $T_l = 300$K (red diamonds), and $T_l = 600$K (purple triangles). Dashed lines in panel (a) mark the three lattice temperatures, at which the respective electrical LTP generation vanishes.}}
	\label{fig:6}
\end{figure}

The second figure of merit we explore is the internal quantum efficiency, that is the number of photons emitted by a single electron propagating through a single layer given by
{\color{black}
\begin{equation}
    \eta^{(2)}\left(T_e\right) = \sum_{\mathbf{q}, j} \frac{\Gamma_{\mathbf{q}, j}^{\mathrm{LTP}}(N_{\mathbf{q}, j}^{\text{ss}}(T_e,T_l),T_e)\xi_{\mathbf{q}, j}}{\nu\left(T_e, T_l\right) n_e \mathrm{A}}, \label{eq:eta2}
\end{equation}
in which $\nu\left(T_e, T_l\right)$ is the drift velocity, related to both the temperature of the lattice $T_l$ and of the electron gas $T_e$ as described in Appendix.~\ref{app:balance}, $n_e$ is the electronic density, and $\mathrm{A}$ is the device in-plane area, which cancels from the final result. 
In Fig.~\ref{fig:6}a we plot $\eta_{\mathbf{q}, j}^{(2)}$ as a function of electronic temperature for a 2nm active layer operating at $T_l = 0$K (green squares), $T_l = 300$K (red diamonds), and $T_l = 600$K (purple triangles). When the electronic temperature grows, more LTPs are generated in the system, even if the growth is sublinear due to the increase of electronic group velocity $\nu(T_e,T_l)$ with the electronic temperature, as electrons spend less time in the active region. At finite $T_l$ the $\eta^{(2)}$ is suppressed near the equilibrum value $T_e = T_l$, where the sum in Eq.~\ref{eq:eta2} vanishes. The asymptotic behaviour $T_e \gg T_l$ is unaffected by the finite temperature of the lattice.
In Fig.~\ref{fig:6}b} we plot the same quantity as a function of the active layer thickness at fixed electronic temperature $T_e = 1000$K for each lattice temperature, showing a saturation of the efficiency for thicker layers in which nonlocal effects decrease as shown in \Eq{eq:couplinga}.
The maximal quantum efficiency is of the order of $10^{-6}$ per nanometric layer.
Superlattice structures can be used to increase the total emission efficiency, while being much more robust to imperfections and easier to fabricate that those required for quantum cascade devices. 

\section{Conclusions and perspectives}
{In this work we have investigated a novel method of mid-infrared light-generation, exploiting the intrinsic optical nonlocality of polar nano-devices. Utilising longitudinal-transverse polaritons (LTPs) as interconnects between the microscopic degrees of freedom of a polar lattice and the far-field, we have demonstrated a sizeably improved narrow-band emission efficiency is achievable when comparing to equivalent thermal mid-infrared emitters.  We illustrated this using a simple, high-symmetry system in order to provide an unbiased assessment of the fundamental process of electrical LTP generation. {\color{black}An optimised device could utilize more complex geometries, as a channel waveguide or nano-resonator to spatially restrict LTP excitation, and to manipulate generated LTPs, but our aim in this paper is to demonstrate the fundamental efficiency of a novel emission channel and we leave device optimization for future works. We hope our results will open the way toward an experimental observation of LTP-driven electrically excited emission and from there to a novel generation of mid-infrared optoelectronic devices.}

\section*{Funding}
\label{SecAck}
S.D.L. is supported by a Royal Society Research fellowship and the Philip Leverhulme prize. The authors acknowledge support from the Royal Society Grant No. RGF\textbackslash EA\textbackslash181001 and the Leverhulme Trust Grant No. RPG-2019-174.

\appendix

\section{Local and Nonlocal Spectra}
\subsection{Local Response of Thin Polar Films}
\label{app:spectral}
In this work we consider a trilayer waveguide, consisting of a polar film of thickness $d$ sandwiched between semi-infinite positive dielectric cladding layers, occupying the region $-d<z<0$.
The system is translationally invariant in the $xy$ plane. Modes of the system are characterized by a 2D in-plane wavevector $\mathbf{q}$, and out-of-plane wavevectors given by
\begin{align}
	\alpha &= \sqrt{\lvert \mathbf{q} \rvert^2 - \epsilon \omega^2 / c^2},\nonumber \\
	\alpha_{\mathrm{c}} &= \sqrt{\lvert \mathbf{q} \rvert^2 - \epsilon_{\mathrm{c}} \omega^2 / c^2},
\end{align}
respectively in the film and the cladding,
where $\epsilon \; (\epsilon_{\mathrm{c}})$ is the dielectric function in the film (cladding) and $c$ is the speed of light in vacuum. In all the Paper in-plane 2D vectors are bold, and three dimensional vectors are indicated either listing the in- and out-of-plane components in squared parenthesis or, when notationally clear, with an arrow symbol.
The full 3D wavevector of the mode in the layers is thus written as $\left[\mathbf{q}, \alpha\right]$, and similarly we write the full 3D coordinate $\left[\mathbf{r}, z\right]$. 

Each interface between the polar film and cladding supports a surface phonon polariton with electric field components at the upper and lower interfaces given by \cite{Gubbin2022b}
\begin{align}
   \vec{E}_{u, \mathbf{q}} &= \begin{cases}
    \frac{\alpha_{\mathrm{c}} \epsilon}{\alpha \epsilon_{\mathrm{c}}} \left[\frac{\mathbf{q}}{\lvert \mathbf{q} \rvert}, - \frac{i \lvert \mathbf{q} \rvert}{\alpha_{\mathrm{c}}} \right]  e^{\alpha_{\mathrm{c}} \left(z + d\right)} e^{i \mathbf{q} \cdot \mathbf{r}} & z < -d,\\
	- \left[\frac{\mathbf{q}}{\lvert \mathbf{q} \rvert},  \frac{i \lvert \mathbf{q} \rvert}{\alpha}\right] e^{-\alpha \left(z + d\right)} e^{i \mathbf{q} \cdot \mathbf{r}}  & -d < z < 0,\\
	- \frac{\alpha_{\mathrm{c}} \epsilon}{\alpha \epsilon_{\mathrm{c}} } \left[\frac{\mathbf{q}}{\lvert \mathbf{q} \rvert},  \frac{i \lvert \mathbf{q}\rvert}{\alpha_{\mathrm{c}}}\right] e^{-\alpha_{\mathrm{c}} z - \alpha d} e^{i \mathbf{q} \cdot \mathbf{r}}  & z > 0,
	\end{cases}  \nonumber \\
    \vec{E}_{l, \mathbf{q}} &= \begin{cases}
    \frac{\alpha_{\mathrm{c}} \epsilon}{\alpha \epsilon_{\mathrm{c}} } \left[\frac{\mathbf{q}}{\lvert \mathbf{q} \rvert}, - \frac{i \lvert \mathbf{q} \rvert}{\alpha_{\mathrm{c}}} \right] e^{\alpha_{\mathrm{c}} \left(z + d\right) - \alpha d} e^{i \mathbf{q} \cdot \mathbf{r}} & z < -d,\\
	\left[\frac{\mathbf{q}}{\lvert \mathbf{q} \rvert},  -\frac{i \lvert \mathbf{q} \rvert}{\alpha}\right]e^{\alpha z } e^{i \mathbf{q} \cdot \mathbf{r}} & -d < z < 0,\\
	-\frac{\alpha_{\mathrm{c}} \epsilon}{\alpha \epsilon_{\mathrm{c}} } \left[\frac{\mathbf{q}}{\lvert \mathbf{q} \rvert},  \frac{i \lvert \mathbf{q} \rvert}{\alpha_{\mathrm{c}}}\right] e^{-\alpha_{\mathrm{c}} z } e^{i \mathbf{q} \cdot \mathbf{r}} & z > 0,
	\end{cases} \label{eq:sphpl}
\end{align}
respectively. Denoting by $\vec{z}=\left[\mathbf{0}, 1\right]$ the unit vector along $z$, the magnetic field in each case can be found utilizing the Maxwell-Faraday equation as
\begin{align}
    \vec{H}_{u, \mathbf{q}} &= \frac{i \omega \epsilon}{c^2 \alpha} \left[ \vec{z} \times \frac{\mathbf{q}}{\lvert \mathbf{q} \rvert}, 0 \right]  e^{i \mathbf{q} \cdot \mathbf{r}}  \begin{cases}
    e^{\alpha_{\mathrm{c}} \left(z + d\right)}& z < -d,\\
    e^{-\alpha \left(z + d\right)}  & -d < z < 0,\\
	e^{-\alpha_{\mathrm{c}} z - \alpha d}  & z > 0, 
	\end{cases} \nonumber \\
    \vec{H}_{l, \mathbf{q}} &=  \frac{i \omega \epsilon}{c^2 \alpha} \left[ \vec{z} \times \frac{\mathbf{q}}{\lvert \mathbf{q} \rvert}, 0 \right]  e^{i \mathbf{q} \cdot \mathbf{r}} \begin{cases}
    e^{\alpha_{\mathrm{c}} \left(z + d\right) - \alpha d}  & z < -d,\\
	e^{\alpha z } & -d < z < 0,\\
	e^{-\alpha_{\mathrm{c}} z } & z > 0,
	\end{cases}
\end{align}
which have been constructed for continuity at the film interfaces $z=0, -d$.  The modes of the waveguide are linear superpositions of those on each of it's interfaces with electric field
\begin{equation}
    \vec{E}_{\mathbf{q}} = \mathrm{U} \vec{E}_{u, \mathbf{q}} + \mathrm{L} \vec{E}_{l, \mathbf{q}},
\end{equation}
where $\mathrm{U, L}$ are coefficients describing the contribution of the SPhPs at either interface. After applying boundary conditions on the tangential component of the electric field $\vec{E}_{\mathbf{q}} \cdot \mathbf{r}$ we recover the following dispersion relation
\begin{equation}
    \left[\begin{array}{cc}
    (\frac{\alpha_{\mathrm{c}} \epsilon}{\alpha \epsilon_{\mathrm{c}} } - 1) e^{- \alpha d} &  1 + \frac{\alpha_{\mathrm{c}} \epsilon}{\alpha \epsilon_{\mathrm{c}} } \\
    \frac{\alpha_{\mathrm{c}} \epsilon}{\alpha \epsilon_{\mathrm{c}} } + 1 & (\frac{\alpha_{\mathrm{c}} \epsilon}{\alpha \epsilon_{\mathrm{c}} } - 1) e^{- \alpha d}
    \end{array}\right]
    \left(\begin{array}{c}
    \mathrm{U} \\
    \mathrm{L}
    \end{array}\right)
    = \left(\begin{array}{c}
    0 \\
    0
    \end{array}\right). \label{eq:dispmat}
\end{equation}
This equation is satisfied when the determinant of the matrix is zero, which leads to the transcendental dispersion relation
\begin{equation}
     1 - r^2 e^{- 2 \alpha d} = 0, \label{eq:displ}
\end{equation}
in which $r$ is the Fresnel reflection coefficient
\begin{equation}
    r = \frac{\alpha \epsilon_{\mathrm{c}} - \alpha_{\mathrm{c}} \epsilon}{\alpha \epsilon_{\mathrm{c}} + \alpha_{\mathrm{c}} \epsilon}.
\end{equation}
After manipulation \Eq{eq:displ} can be recast in the form \cite{Gubbin2022b}
\begin{equation}
	2 - \tanh\left(\alpha d\right) \left[ \frac{\epsilon \alpha_\mathrm{c}}{\epsilon_\mathrm{c} \alpha} + \frac{\epsilon_\mathrm{c} \alpha}{\epsilon \alpha_\mathrm{c}}\right] = 0. \label{eq:locdisp}
\end{equation}
Although in the thin film limit \Eq{eq:locdisp} can simplify dramatically \cite{Campione2015} in this work we cannot utilise this approximation  as it breaks down at the larger wavevectors under study. The SPhP frequency $\omega_{\mathbf{q}}^{\mathrm{SP}}$ utilised in the main body is found by numerical solution of \Eq{eq:locdisp}. This yields a complex modal frequency as a function of wavevector $\mathbf{q}$, which leads to the loss rate
\begin{equation}
   \gamma^M_{\mathbf{q}}=  \Im \left\{ \omega_{\mathbf{q}}^{\mathrm{SP}} \right\} = \frac{\gamma}{2},
\end{equation}
in the wavevector region under study \cite{Campione2015}, a consequence of strong confinement in the polar layer. Note that we neglect the frequency and polarization dependence of the phononic loss rate.

\subsection{Nonlocal Response of Thin Polar Films}
\label{app:spectranl}
The nonlocal optics of the system  can be derived semi-analytically considering the dispersion of LO phonons in the polar film \cite{Gubbin2022b}. In the typical quadratic model of phonon dispersion \cite{Gubbin2020} the dispersive LO phonon frequency takes the form
\begin{equation}
	\omega^{\mathrm{L}\,2}_{\vec{q}} = \omega_{\mathrm{L}}^2 - \beta_{\mathrm{L}}^2 \left[\lvert \mathbf{q} \rvert^2 + q_z^2\right],
\end{equation}
where $\omega_{\mathrm{L}}$ is the frequency at zero-wavevector, $\vec{q}=\left[\mathbf{q}, q_z\right]$ is the 3D wavevector and $\beta_{\mathrm{L}}$ is a characteristic velocity. In a planar layer enforcing Pekar boundary conditions, as when the Reststrahlen bands in the layer and the cladding do not overlap, this leads to discrete, localised LO phonon modes with frequencies
\begin{equation}
	\omega_{\mathbf{q},n}^{\mathrm{L}} = \sqrt{\omega_{\mathrm{L}}^2 - \beta_{\mathrm{L}}^2 \left(\xi_n^2 + \lvert \mathbf{q} \rvert^2\right)}, \label{eq:phononfreq}
\end{equation}
where $\xi_n$ is the quantized out-of-plane wavevector defined in \Eq{eq:LOwv} in the main text.\\
The SPhP mode considered in Appendix~\ref{app:spectral} and these localised LO phonons interact through shared electro-mechanical boundary conditions \cite{Gubbin2020}. In the trilayer under study we previously derived the analytical form of this coupling $\Omega_{n, \mathbf{q} }$, given by \Eq{eq:couplinga} \cite{Gubbin2022b}. Using this result the interacting system can be described in the rotating wave approximation by the quadratic Hamiltonian 
\begin{align}
	\HLTP =& \hbar\sum_{\mathbf{q} } \biggr \{   \freqn{SP}{\mathbf{q}} \crsp{\mathbf{q} } \ansp{\mathbf{q} } 
	 + \sum_{n } \biggr[ \freqn{L}{ \mathbf{q},n } \crlo{ \mathbf{q},n } \anlo{ \mathbf{q},n } \nonumber
	\\ &  \quad \quad \quad  + \Omega_{ \mathbf{q},n } \left(\ansp{\mathbf{q} }\crlo{\mathbf{q},n}+\crsp{\mathbf{q}}\anlo{\mathbf{q},n}  \right)\biggr]\biggr\}, \label{eq:HLTP}
\end{align}
in which $\crsp{\mathbf{q}} \; (\crlo{\mathbf{q}, n})$ are the annihilation operators for the SPhP ($n^{\text{th}}$ LO phonon) mode, obeying bosonic commutation rules
\begin{align}
 \left[\ansp{\mathbf{q}},\crsp{\mathbf{q}'} \right]&=\delta_{\mathbf{q},\mathbf{q}'},\nonumber\\ 
 \left[\anlo{\mathbf{q},n},\crlo{\mathbf{q}',n'} \right]&=\delta_{\mathbf{q},\mathbf{q}'}\delta_{n,n'}.
\end{align}
The quadratic bosonic Hamiltonian in \Eq{eq:HLTP} can be diagonalized using a multimode Hopfield-Bogoliubov diagonalization into a spectrum of longitudinal-transverse polaritons (LTPs) \cite{Gubbin2022b} with frequencies $\omega^{\text{LTP}}_{\mathbf{q}, j}$
\begin{align}
    \HLTP =& \hbar\sum_{\mathbf{q},j } \omega^{\text{LTP}}_{\mathbf{q}, j}\hat{d}_{\mathbf{q},j}^{\dagger}\hat{d}_{\mathbf{q},j},
\end{align}
described by creation operators
\begin{align}
\label{eq:dLTP}
\hat{d}_{\mathbf{q},j}=\alpha_{\mathbf{q},j}\hat{a}_{\mathbf{q}}
+\sum_n \beta_{\mathbf{q},j,n}\hat{b}_{\mathbf{q},n},
\end{align}
in which $j$ indexes the polariton branch, and the Hopfield coefficients $\alpha_{\mathbf{q},j}, \beta_{\mathbf{q},j,n}$ preserve the unitary nature of the transformation
\begin{align}
\lvert\alpha_{\mathbf{q},j}\rvert^2+\sum_n  \lvert\beta_{\mathbf{q},j,n}\rvert^2&=1.
\end{align}
This ensures that the resulting LTP modes are also bosonic, whose creation and annihilation operators satisfy the commutation relation
\begin{align}
\left[\hat{d}_{\mathbf{q},j},\hat{d}_{\mathbf{q}',j'}^{\dagger}\right]=\delta_{\mathbf{q},\mathbf{q}'}\delta_{j,j'},
\end{align}
and $j$ indexes the different LTP branches.\\
The homogeneous lineshape of the $j^{\text{th}}$ LTP branch can be written as a Lorentzian
\begin{align}
\rho_{\mathbf{q}, j}^{\text{LTP}}(\omega)&=\frac{1}{\pi}\frac{\gamma^{\text{LTP}}_{\mathbf{q}, j}}{(\omega-\omega^{\text{LTP}}_{\mathbf{q}, j})^2+(\gamma^{\text{LTP}}_{\mathbf{q}, j})^2},    \label{eq:dos} 
\end{align}
where the homogeneous LTP linewidth can be written from \Eq{eq:dLTP} as a weighed sum of the different involved scattering channels as
\begin{align}
\gamma^{\text{LTP}}_{\mathbf{q}, j}&=\gamma+\lvert \alpha_{\mathbf{q}, j} \rvert^2\left(\gamma^M_{\mathbf{q}}-\gamma \right).
\label{eq:gammaLTP}
\end{align}

\subsection{Quantization of the LO Mode}
\label{app:quant}
In this Paper we aim to describe interactions between electrons and localised LO phonon modes through the Fr{\"o}hlich interaction. To do this we require the Hamiltonian modeling the interaction of the two systems. { 
Given an electronic charge density operator $\hat{\rho}$ we can write
\begin{equation}
	\mathcal{\hat{H}_{\text{Int}}} = - \int \mathrm{d^2 r}\, \mathrm{d z}\, \left[\hat{\rho}\left(\mathbf{r}, z\right) \hat{\phi}\left(\mathbf{r}, z\right) + \hat{\mathbf{j}}\left(\mathbf{r}, z\right) \cdot \hat{\mathbf{A}}\left(\mathbf{r}, z\right) \right], \label{eq:hamelb}
\end{equation}
 where $\hat{\phi}$ is the electric potential arising from LO phonon modes, $\hat{\mathbf{j}}$ is the electric current density operator and $\hat{\mathbf{A}}$ is the electromagnetic vector potential. The first term of this Hamiltonian describes the interaction between an electronic charge and the induced LTP charge in the lattice, while the second describes interaction between electronic currents and the electromagnetic tranverse fields. The former one is dominant for low-frequency currents and in this Paper we will consider only this component, using the Fr{\"o}hlich Hamiltonian
 \begin{equation}
	\mathcal{\hat{H}}_{\mathrm{Fr\ddot{o}}} = - \int \mathrm{d^2 r}\, \mathrm{d z}\, \hat{\rho}\left(\mathbf{r}, z\right) \hat{\phi}\left(\mathbf{r}, z\right) . \label{eq:hamel}
\end{equation}
 In a local model the discontinuity in the out-of-plane electric field at the boundaries of the polar layer leads to a finite induced charge which, due to the transverse nature of the electric field, is unable to spread into the interior of the polar layer. In our nonlocal model instead, charge induced at the edge of the polar layer is able to spread into the interior, manifesting as a source of LO phonons. This charging is entirely encoded in the LO phonon fields and their corresponding scalar potential.
 }

The electric potential at fixed in-plane wavevector $\mathbf{q}$ in branch $n$ with out-of-plane wavevector $\xi_n$ given by \Eq{eq:LOwv} is proportional to
\begin{equation}
	u_{\mathbf{q}, n}\left(\mathbf{r}, z\right) = \sin\left[\xi_n \left(z+d/2\right)\right] e^{i \mathbf{q} \cdot  \mathbf{r}}.
	\label{eq:phonwave}
\end{equation}
To parameterize the Hamiltonian \Eq{eq:hamel} it is necessary to quantize the phonon field. This can be carried out from the corresponding electric field
\begin{equation}
	\vec{E}_{\mathbf{q}, n} \left(\mathbf{r}, z\right)  = - \mathrm{B}_{\mathbf{q}, n} \nabla u_{\mathbf{q}, n}\left(\mathbf{r}, z\right),
\end{equation}
where $ \mathrm{B}_{\mathbf{q}, n}$ is a constant to be determined by the quantization procedure. Utilizing the results of Refs. \cite{Gubbin2016b, Gubbin2022c}, where we developed a quantum theory on local and nonlocal polar media,
\begin{align}
	\frac{2 \epsilon_0}{\hbar \omega_{\mathbf{q}, n}^{\mathrm{L}}}  \int \mathrm{d^3 r} \frac{\epsilon_{\infty} \omega_{\mathbf{q}, n}^{\mathrm{L} \; 2} \left[\omega_{\mathrm{L}}^2 - \omega_{\mathrm{T}}^2\right]}{\left[ \omega_{\mathbf{q}, n}^{\mathrm{L} \; 2}  -  \omega_{\mathrm{T}}^2  + \beta_{\mathrm{L}}^2 \left(\lvert \mathbf{q}\rvert^2 + \xi_n^2\right)\right]^2} \lvert \vec{E}_{\mathbf{q}, n}\rvert^2 
&	\nonumber \\
=	\mathrm{sgn}\left(\omega_{\mathrm{q}, n}^{\mathrm{L}}\right),& \label{eq:LOq1}
\end{align}
where $\omega_{\mathbf{q}, n}^{\mathrm{L}}$ is the mode frequency, given in this case be \Eq{eq:LOFreq}. For the Fabry-P{\'e}rot modes under study, considering that $\xi_n = n \pi / d$ we can find
\begin{align}
    \int \mathrm{d^3 r}  \lvert \vec{E}_{\mathbf{q}, n}\rvert^2 = \lvert \mathrm{B}_{\mathbf{q}, n} \rvert^2 \frac{ \mathrm{A} d}{2 \beta_{\mathrm{L}}^2} \left[\omega_{\mathrm{L}}^2 - \omega_{\mathbf{q}, n}^2 \right],
    \label{eq:x}
\end{align}
with $A$ the device in-plane area.
After substituting \Eq{eq:x} back into \Eq{eq:LOq1} we recover the quantization constant 
\begin{align}
    \mathrm{B}_{\mathbf{q}, n} &=\sqrt{\frac{\hbar \omega_{\mathbf{q}, n}^{\mathrm{L}}}{\epsilon_0 \epsilon_{\mathbf{q}, n}^{\rho} \mathrm{A} d}} \sqrt{\frac{\beta_{\mathrm{L}}^2}{\omega_{\mathrm{L}}^2 - \omega_{\mathbf{q}, n}^{\mathrm{L} \; 2}}}, \label{eq:Bfac}
\end{align}
in which, compared to the local case $\omega_{\mathbf{q}, n}^{\mathrm{L}} = \omega_{\mathrm{L}}$, we acquired a dispersive Fr{\"o}hlich constant
\begin{equation}
    \epsilon_{\mathbf{q}, n}^{\rho} = \epsilon_{\infty} \frac{\omega_{\mathbf{q}, n}^{\mathrm{L} \; 2}}{\omega_{\mathrm{L}}^2 - \omega_{\mathrm{T}}^2}.
\end{equation}
Finally, we can write the second-quantized form of the LO phonon potential
\begin{equation}
	\hat{\phi}\left(\mathbf{r}, z\right) = \sum_{\mathbf{q}, n} \mathrm{B}_{\mathbf{q}, n} u_{\mathbf{q}, n}\left(\mathbf{r}, z\right) \left[\anlo{\mathbf{q}, n} + \crlo{-\mathbf{q}, n}\right],
\end{equation}
which parameterizes the LO phonon contribution to \Eq{eq:hamel}.

\section{Electron Gas}
\label{app:eg}
Electrons interact with LO phonons through the Fr\"ohlich Hamiltonian in \Eq{eq:hamel}. Introducing the fermionic annihilation operators $\hat{c}_{\vec{k}}$ with 3D momentum $\vec{k}=\left[\mathbf{k}, k\right]$, describing electrons in the conduction band, the free electron Hamiltonian is given by
\begin{equation}
\label{eq:fel}
	\hat{\mathcal{H}} = \sum_{\vec{k}}  \varepsilon_{\vec{k}} \hat{c}_{\vec{k}}^{\dagger} \hat{c}_{\vec{k}},
\end{equation}
where the free-electron energy is $ \varepsilon_{\vec{k}} = \hbar^2 \lvert \vec{k} \rvert^2 / 2 m^*$ and $m^*$ is the effective electron mass in the cladding region. We consider a Si cladding region and we will thus consider the conductivity effective mass $m^*=0.26m_e$, where $m_e$ is the free-electron mass. 
The electronic creation and annihilation operators obey Fermionic anti-commutator rules
\begin{align}
\{ \hat{c}_{\vec{k}},\hat{c}_{\vec{k}'}^{\dagger}\}
&=\delta_{\vec{k},\vec{k'}},
\end{align}
and have free plane-wave wavefunctions of form
\begin{equation}
    \psi_{\vec{k}} \left(\mathbf{r}, z\right) = \frac{1}{\sqrt{\mathrm{L A}}} e^{i \mathbf{k} \cdot \mathbf{r}} e^{i k z},
\end{equation}
where $\mathrm{L}$ is the electronic quantization length along $z$ and $\mathrm{A}$ the sample surface. By inverting \Eq{eq:dLTP} to write the LO phonon operators in terms of the LTPs we can rewrite \Eq{eq:hamel} as
\begin{align}
	\hat{\mathcal{H}}_{\mathrm{Fr\ddot{o}}} =  \frac{(2 \pi)^2}{\mathrm{A}}
	\sum_{\vec{k},\mathbf{q}, \vec{k}', j} \hbar\kappa_{\mathbf{q}, k, k', j} &\delta\left(\mathbf{k' - k - q}\right) \nonumber \\
	&\quad \times 
    \left[ 
	\hat{c}_{\vec{k}}^{\dagger}\hat{c}_{\vec{k}'} \hat{d}^{\dagger}_{\mathbf{q}, j} 
	+\hat{c}_{\vec{k}'}^{\dagger}\hat{c}_{\vec{k}} \hat{d}_{\mathbf{q},j } \right], \label{eq:HF2nd}
\end{align}
in which $\vec{k} = \left[\mathbf{k}, k\right]$ is the incoming electronic wavevector and $\vec{k}' = \left[\mathbf{k}', k'\right]$ is the outgoing one.  Integrating over the in-plane component of the out-going wavevector $\mathbf{k}'$ we put the Fr{\"o}hlich Hamiltonian in the form
\begin{align}
	\hat{\mathcal{H}}_{\mathrm{Fr\ddot{o}}} = & 
	\sum_{\vec{k}, k', \mathbf{q}, j} \hbar\kappa_{\mathbf{q}, k, k',j }
    \left[ 
	\hat{c}_{\vec{k}}^{\dagger}\hat{c}_{\vec{k} + \vec{q}} \hat{d}^{\dagger}_{\mathbf{q}, j} 
	+\hat{c}_{\vec{k} + \vec{q}}^{\dagger}\hat{c}_{\vec{k}} \hat{d}_{\mathbf{q}, j} \right],
\end{align}
clearly describing the dressed Fr\"ohlich process in Fig.~\ref{fig:2}c,
where the electron-LTP coupling coefficient has form
\begin{align}
\label{eq:kappaF}
\kappa_{\mathbf{q}, k, k', j}&= \frac{e}{\hbar} \sum_n \mathrm{B}_{\mathbf{q}, n}\beta_{\mathbf{q}, j,n}\Xi_{k,k', n},
\end{align}
with the electron-phonon overlap integral given by
\begin{align}
\Xi_{k, k', n} = \frac{1}{\mathrm{L}}  \int_{-d}^0 \mathrm{d z}\, e^{i \left(k - k'\right) z} \sin\left[\xi_n \left(z + d / 2\right)\right]. \label{eq:Xi}
\end{align}

%
%
%
Note that in systems containing free charges the dielectric function differs from \Eq{eq:Lorentz}, and is instead given by
\begin{equation}
    \epsilon\left(\omega\right) = \epsilon_{\infty} \frac{\omega_{\mathrm{L}}^2 - \omega \left(\omega + i \gamma\right)}{\omega_{\mathrm{T}}^2 - \omega \left(\omega + i \gamma\right)} - \sum_{j = e, h} \frac{n_j e^2}{\epsilon_0 m_j^*} \frac{1}{\omega\left(\omega + i \Gamma\right)},
\end{equation}
where $\Gamma$ is a characteristic free carrier scattering rate, {$n_{j}$ is the carrier density, $m_j^*$ is the carrier conductivity effective mass, and $j$ refers to electrons $e$, or holes $h$}. In SiC the relatively high frequencies of the Reststrahlen band mean that even at large charge densities the shift in dielectric function is minimal \cite{Dunkelberger2018}. Deviations are most severe near to $\omega_{\mathrm{L}}$ where the Lorentz dielectric function vanishes. Note that the predominant effect is a small blue shift in SPhP frequencies. This would actually be beneficial for the effect we aim to describe, as the SPhP mode would move toward $\omega_{\mathrm{L}}$ where electron-phonon coupling is strongest.\\

\section{Thermalized Electron Model}
\label{app:balance}
Electrons interacting with the polar nanolayer considered in this Paper are assumed to propagate under the influence of an applied electric field polarized perpendicular to the nanolayer. An average electron propagates at the \emph{drift velocity}, which is limited through momentum relaxation with the crystal lattice. The lattice also provides energy relaxation channels, which reduce the mean electronic energy and prevent the electron gas from overheating. We describe the electron gas as thermalised at a certain electronic temperature $T_e$.
In the following we calculate closed relationships between the applied field, the drift velocity, and the temperature of the electron gas.\\
We are able to write the balance equations between electronic energy $\varepsilon$ and drift velocity $\nu$ as \cite{khurgin_hot_2007}
\begin{subequations}
\begin{align}
    \frac{\partial \varepsilon}{\partial t} &= - e \mathcal{E} \nu - \frac{\hbar \omega_{\mathrm{L}}}{\tau_{\varepsilon} \left(T_e,T_l\right)}, \label{eq:eom}\\
    \frac{\partial \nu}{\partial t} &= - \frac{e \mathcal{E}}{m^*} - \frac{\nu}{\tau_m\left(T_e,T_l\right)},
\end{align}
\end{subequations}
where $\mathcal{E}$ is the scalar electric field, $\tau_{\varepsilon}\left(T_e,T_l\right) \; (\tau_m\left(T_e,T_l\right))$ is the energy (momentum) relaxation time dependent on both the electron and lattice temperatures. In the steady state these equations can be directly solved
\begin{align}
    e \mathcal{E} \nu &= - \frac{\hbar \omega_{\mathrm{L}}}{\tau_{\varepsilon} \left(T_e,T_l\right)}, \\
    \frac{e \mathcal{E}}{m^*} &=  - \frac{\nu}{\tau_m\left(T_e,T_l\right)},
\end{align}
allowing the electric field or drift velocity to be eliminated
\begin{subequations}
\begin{align}
    m^* \nu^2 &= \hbar \omega_{\mathrm{L}} \frac{\tau_m\left(T_e,T_l\right)}{\tau_{\varepsilon} \left(T_e,T_l\right)},\\
    \frac{e^2 \mathcal{E}^2}{m^*}  &= \frac{\hbar \omega_{\mathrm{L}}}{\tau_m\left(T_e,T_l\right) \tau_{\varepsilon} \left(T_e,T_l\right)},
\end{align}
\end{subequations}
Although it is not strictly possible to define a relaxation time for the polar interaction, a consequence of the inelastic and non velocity-randomising nature of the interaction \cite{jacoboni_theory_2010}, it is possible to approximate an energy and momentum relaxation time utilizing Fermi's golden rule. Following Conwell \cite{conwell_high_1969}, the probability per unit time of a carrier scattering out of state with momentum $\vec{k} = \left[\mathbf{k}, k\right]$ with exchanged momentum $\vec{q} = \left[\mathbf{q}, q\right]$ is given by
\begin{widetext}
\begin{align}
    \frac{1}{\tau_{\vec{k}}(T_l)} = & \frac{2 \pi}{\hbar} \sum_{\vec{q}} 
     \biggr[ \lvert \langle \vec{k} + \vec{q}, N_\mathrm{L} - 1 \lvert \hat{\mathcal{H}}_{\mathrm{Fr\ddot{o}}} \rvert \vec{k}, N_\mathrm{L}\rangle \rvert^2 \delta \left(\varepsilon_{\vec{k} + \vec{q}} - \varepsilon_{\vec{k}} - \hbar \omega_{\mathrm{L}}\right) 
     + \lvert \langle \vec{k} - \vec{q}, N_\mathrm{L} + 1 \lvert \hat{\mathcal{H}}_{\mathrm{Fr\ddot{o}}} \rvert \vec{k}, N_\mathrm{L}\rangle \rvert^2 \delta \left(\varepsilon_{\vec{k} - \vec{q}} - \varepsilon_{\vec{k}} + \hbar \omega_{\mathrm{L}}\right) \biggr],
\end{align}
\end{widetext}
in which the first term is the transition probability from phonon absorption, the second is from phonon emission. Here $N_\mathrm{L}\left(T_l\right)$ is the equilibrium phonon population in the LO phonon modes which are assumed to be non-dispersive. If we assume spherical bands we can compute the inverse scattering time directly as
\begin{multline}
    \frac{1}{\tau_{\vec{k}}\left(T_l\right)} = \frac{2 e \mathcal{E}_{\textrm{F}}}{\sqrt{2 m^* \varepsilon_{\vec{k}}}} \biggr[ N_{\mathrm{L}}\left(T_l\right) \sinh^{-1} \left(\sqrt{\frac{\varepsilon_{\vec{k}}}{\hbar \omega_{\mathrm{L}}}}\right) \\
    + \left(N_{\mathrm{L}}\left(T_l\right) + 1 \right) \sinh^{-1} \left(\sqrt{\frac{\varepsilon_{\vec{k}}}{\hbar \omega_{\mathrm{L}}} - 1}\right) \biggr],
\end{multline}
where $\mathcal{E}_{\textrm{F}}$ is the bulk Fr{\"o}hlich matrix element
\begin{equation}
    e \mathcal{E}_{\textrm{F}} = \frac{m^* e^2 \hbar \omega_{\mathrm{L}}}{4 \pi \hbar^2 \epsilon_{0}} \left[\frac{1}{\epsilon_{\infty}} - \frac{1}{\epsilon_{\text{st}}}\right],
\end{equation}
in which $\epsilon_{\infty} \; (\epsilon_{\text{st}})$ is the high-frequency (static) dielectric constant of the lattice and $\epsilon_{0}$ is the permittivity of free-space.\\
The inverse scattering time $\tau_{\vec{k}}^{-1}\left(T_l\right)$ is not a relaxation time, and as discussed above one cannot be strictly defined for the polar interaction. We can however define the rate of energy change for an electron with momentum $\vec{k}$ due to the polar interaction, given by 
\begin{equation}
  \left.\frac{\mathrm{d} \varepsilon_{\vec{k}}}{\mathrm{d} t} \right|_\mathrm{PI} = \frac{\hbar \omega_\mathrm{L}}{\tau_{\vec{k}}\left(T_l\right)},
\end{equation}
 Pseudo-relaxation times can then be calculated by taking the distributional average of this rate of energy change. Assuming the electron gas is described by a Maxwell-Boltzmann distribution at electronic temperature $T_e$
\begin{equation}
  f_{\vec{k}}\left(T_e\right) = \frac{8\pi^3}{LA} \frac{ e^{- \varepsilon_{\vec{k}} / k_{\mathrm{B}} T_e}}{N_\mathrm{c} \left(T_e\right)}, \label{eq:f}
\end{equation}
where $N_\mathrm{c}\left(T_e\right) = 2 \eta \left[\frac{m^* k_{\mathrm{B}} T_e}{2 \pi \hbar^2}\right]^{3/2}$ is the density of states in the conduction band, $n_e$ is the electronic density  and $\eta$ is the degeneracy factor in the conduction band.
Integrating over the distribution we find the energy pseudo-relaxation time 
\begin{widetext}
\begin{align}
    \frac{1}{\tau_{\varepsilon}\left(T_e, T_l\right)} &\approx \frac{1}{\hbar \omega_{\mathrm{L}}}\biggr\langle\left. \frac{\mathrm{d} \varepsilon_{\vec{k}}}{\mathrm{d} t}\right|_{\mathrm{PI}} \biggr\rangle
    = - \sqrt{\frac{2 }{\pi m^*\hbar \omega_\mathrm{L}}} e \mathcal{E}_{\textrm{F}} \frac{e^{g\left(T_l\right) - g\left(T_e\right)/2} - e^{g\left(T_e\right) / 2}}{e^{g\left(T_l\right)} - 1}  \sqrt{g\left(T_e\right)}  K_0\left(\frac{g\left(T_e\right)}{2}\right),
\end{align}
where the angular brackets are an average calculated over the distribution 
in \Eq{eq:f}, the $K_i$ are modified Bessel functions, and we defined the dimensionless quantity $g\left(T\right) = \hbar \omega_{\mathrm{L}} / k_{\mathrm{B}}T $. \\
To find the momentum pseudo-relaxation time we start instead by calculating the change of electronic wavevector in the transport direction $\vec{z}$ for a carrier with momentum $\vec{k} = \left[\mathbf{k}, k\right]$, given by 
\begin{align}
    \frac{\mathrm{d} k}{\mathrm{d} t} &= \frac{2 \pi}{\hbar} \sum_{\vec{q}} 
  \biggr[ k \lvert \langle \vec{k} + \vec{q}, N_{\mathrm{L}} - 1 \lvert \mathcal{H}_{\mathrm{Fr\ddot{o}}} \rvert \vec{k}, N_{\mathrm{L}}\rangle \rvert^2 \delta \left(\varepsilon_{\vec{k} + \vec{q}} - \varepsilon_{\vec{k}} - \hbar \omega_{\mathrm{L}}\right)    - k \lvert \langle \vec{k} - \vec{q}, N_{\mathrm{L}} + 1 \lvert \mathcal{H}_{\mathrm{Fr\ddot{o}}} \rvert \vec{k}, N_{\mathrm{L}}\rangle \rvert^2 \delta \left(\varepsilon_{\vec{k} - \vec{q}} - \varepsilon_{\vec{k}} + \hbar \omega_{\mathrm{L}}\right) \biggr].
\end{align}
As for the previous case we can integrate analytically, yielding the result
\begin{multline}
    \frac{\mathrm{d} k}{\mathrm{d} t} 
    = - \frac{e \mathcal{E}_{\textrm{F}}}{\hbar} \frac{k}{\lvert \vec{k}\rvert} \biggr[ N_{\mathrm{L}} \biggr \{\sqrt{1 + \frac{\hbar \omega_{\mathrm{L}}}{\varepsilon_{\vec{k}}}} - \frac{\hbar \omega_{\mathrm{L}}}{\varepsilon_{\vec{k}}} \sinh^{-1} \left(\sqrt{\frac{\varepsilon_{\vec{k}}}{\hbar \omega_{\mathrm{L}}}}\right)  \biggr\} 
    + (N_{\mathrm{L}} + 1) \biggr \{\sqrt{1 - \frac{\hbar \omega_{\mathrm{L}}}{\varepsilon_{\vec{k}}}} + \frac{\hbar \omega_{\mathrm{L}}}{\varepsilon_{\vec{k}}} \sinh^{-1} \left(\sqrt{1-\frac{\hbar \omega_{\mathrm{L}}}{\varepsilon_{\vec{k}}}}\right) \biggr\}\biggr].
\end{multline}
Analogously to the energy relaxation rate we can find the pseudo-relaxation time by integrating over the electronic distribution. To do this we need to assume the distribution function is drifted, as the gas propagates at drift velocity $\nu$ along $\vec{z}$. The zero-order spherically symmetric term does not contribute, the first-order perturbation of the distribution is
\begin{equation}  
    f^{(1)}_{\vec{k}}\left(T_e\right) = k \frac{\hbar \nu}{k_{\mathrm{B}} T_e} \frac{8\pi^3}{LA}\frac{e^{-\varepsilon_{\vec{k}} / k_{\mathrm{B}} T_e}}{N_\mathrm{c}\left(T_e\right)},
\end{equation}
and integrating over the distribution yields the momentum pseudo-relaxation time 
\begin{align}
    \frac{1}{\tau_m\left(T_e, T_l\right)} \approx&  \frac{1}{m^* \nu} \biggr \langle \left.\frac{\mathrm{d} \left( \hbar k\right)}{\mathrm{d} t}\right|_{\textrm{PI}} \biggr \rangle = - \frac{2 e \mathcal{E}_{\textrm{F}} N_{\mathrm{L}} \left(T_l\right)}{3 \sqrt{\pi}} \frac{g\left(T_e\right)^{3/2}}{\sqrt{2 m^* \hbar \omega_{\mathrm{L}}}} e^{g\left(T_e\right)/2}\nonumber \\ &\times \biggr[\left(e^{g\left(T_l\right) - g\left(T_e\right)}+1\right) K_1\left(\frac{g\left(T_e\right)}{2}\right) + \left(e^{g\left(T_l\right) - g\left(T_e\right)}-1\right) K_0\left(\frac{g\left(T_e\right)}{2}\right)\biggr].
\end{align}
We thus arrive at our results for the drift velocity and the field intensity
\begin{subequations}
\begin{align}
    \nu^2 &= \frac{\hbar \omega_{\mathrm{L}}}{m^*} \frac{\tau_m\left(T_e, T_l\right)}{\tau_{\varepsilon} \left(T_e, T_l\right)}= \frac{3 \hbar \omega_{\mathrm{L}}}{\sqrt{2} m^* g\left(T_e\right)} \frac{\left[e^{g\left(T_l\right) - g\left(T_e\right)} - 1\right]  K_0\left(\frac{g\left(T_e\right)}{2}\right)}{ \biggr[\left(e^{g\left(T_l\right) - g\left(T_e\right)}+1\right) K_1\left(\frac{g\left(T_e\right)}{2}\right) + \left(e^{g\left(T_l\right) - g\left(T_e\right)}-1\right) K_0\left(\frac{g\left(T_e\right)}{2}\right)\biggr]},\\
    \left[\frac{\mathcal{E}\left(T_e, T_l\right)}{\mathcal{E}_{\textrm{F}}}\right]^2 =& \frac{m^* \hbar \omega_{\mathrm{L}}}{e^2 \mathcal{E}_{\textrm{F}}^2 \tau_m\left(T_e,T_l\right) \tau_{\varepsilon} \left(T_e,T_l\right)} 
    = \frac{2 N_{\mathrm{L}} \left(T_l\right)^2 g\left(T_e\right)^{2}}{3 \pi} e^{g\left(T_e\right)}\left[e^{g\left(T_l\right) - g\left(T_e\right)} - 1\right]  K_0\left(\frac{g\left(T_e\right)}{2}\right)\nonumber \\
    & \times \biggr[\left(e^{g\left(T_l\right) - g\left(T_e\right)}+1\right) K_1\left(\frac{g\left(T_e\right)}{2}\right) + \left(e^{g\left(T_l\right) - g\left(T_e\right)}-1\right) K_0\left(\frac{g\left(T_e\right)}{2}\right)\biggr]. 
\end{align}
\end{subequations}
\end{widetext}

\section{Calculation of Electrical Injection}
\label{app:Thermal}
In this Appendix we calculate the LTP generation rate by a thermalized, non-degenerate electron gas, described by the distribution in \Eq{eq:f}. As described in Appendix~\ref{app:balance} this electronic temperature can be related to the electronic drift velocity. \\
The rate at which electrons with wavevector $\vec{k}$ emit an LTP of in-plane wavevector $\mathbf{q}$ in branch $j$, while exchanging out-of-plane momentum $k'$ is given by
\begin{widetext}

\begin{align}
\Gamma_{\mathbf{q}, \vec{k}, k' ,j}^{\mp} &=2\pi\hbar  n_e\mathrm{A L}
\lvert \kappa_{\mathbf{q}, k, k', j} \rvert^2   \delta(\varepsilon_{\mathbf{k}, k}-\varepsilon_{\mathbf{k \mp q}, k'} \mp \hbar\omega^{\text{LTP}}_{\mathbf{q},j}) f_{\vec{k}}(T_e),
\end{align}
where the $\mp$ refers to emission ($+$) and  absorption ($-$) of a LTP respectively.
Integrating over the incident electronic wavevector $\vec{k}$ we derive the total rate at which all electrons emit LTPs in branch $j$ with in-plane wavevector $\mathbf{q}$
\begin{align}
	\Gamma_{\mathbf{q}, j}^{\mp} &= \sum_{k', \vec{k}} \Gamma_{\mathbf{q}, \vec{k}, k', j}^{\mp} =  \frac{\mathrm{A L}^2}{(2 \pi)^4}  \int \mathrm{d} k' \int \mathrm{d } k\int \mathrm{d}^2 \mathbf{k}\, \Gamma_{\mathbf{q}, \vec{k}, k', j}^{\mp} =  \hbar \mathrm{A L}^2 \frac{n_e}{N_{\mathrm{c}}} \int \mathrm{d} k' \int \mathrm{d } k\int \mathrm{d}^2 \mathbf{k} \lvert \kappa_{\mathbf{q}, k, k', j} \rvert^2   \nonumber \\
	&\quad \times e^{-\frac{\varepsilon_{\vec{k}}}{k_B T_e}} \delta(\varepsilon_{\mathbf{k}, k}-\varepsilon_{\mathbf{k \mp q}, k'} \mp \hbar\omega^{\text{LTP}}_{\mathbf{q},j}).
\end{align}
We account for linewidth broadening of the LTP modes by using, instead of the Dirac's delta, the Lorentzian density of states from \Eq{eq:dos}.
\end{widetext}

\bibliography{bibliography}

\end{document}